\documentclass[11pt]{article}
\usepackage{amsmath}
\usepackage{amsfonts}
\usepackage{amssymb}
\usepackage{graphicx}
\usepackage{bm}
\usepackage{color}
\usepackage{amscd}

\def\bea{\begin{eqnarray}}
\def\eea{\end{eqnarray}}

\begin{document}
\begin{center}
\LARGE {\bf Tachyon-Warm Intermediate and Logamediate Inflation in the Brane-World Model in the Light of Planck Data}
\end{center}
\begin{center}
{V. Kamali $^{a}$\footnote{E-mail: vkamali1362@gmail.com, vkamali@basu.ac.ir}\hspace{1.5mm}
M. R. Setare $^{b}$\footnote{E-mail: rezakord@ipm.ir
}\hspace{1mm} ,
 \\
 $^a$ {\small {\em  Department of Physics, Faculty of Science,\\
Bu-Ali Sina University, Hamedan, 65178, Iran}}}\\
 $^b$ {\small {\em Department of Science, Campus of Bijar, University of Kurdistan.\\
  Bijar , Iran.}}\hspace{1.5mm}\\

\end{center}


\begin{center}
{\bf{Abstract}}\\
Tachyon inflationary universe model on the brane  in the context
of warm inflation is studied. In
slow-roll approximation and in longitudinal gauge, we find the primoradial perturbation spectrums
for this scenario.  We also present the general
expressions of the tensor-scalar ratio, scalar spectral index
and its running. We develop our model by using exponential
potential, the characteristics of this model are calculated in
great details. We also study our model in the context of intermediate (where scale factor expands as: $a=a_0\exp(At^f)$) and
logamediate (where the scale factor expands as: $a=a_0\exp(A[\ln t]^{\nu})$) models of inflation. In these two sectors,  dissipative parameter is considered as a constant parameter and a function of tachyon field. Our model is compatible with observational data.
The parameters of the model are restricted by
Planck data.
 \end{center}

\newpage

\section{Introduction}
Inflation as a theoretical framework  presents the better description of the early phase of our universe.
Main problems of Big Bang model (horizon,
flatness,...) could be solved in the context of inflation scenario \cite{1-i}.
Lagrangian formalism in terms of scalar fields, can explain this scenario.
Quantum fluctuations of the scalar field provide a description of anisotropy of cosmic microwave background (CMB) and origin of the distribution of
large scale structure (LSS) \cite{6,planck}. Standard model of inflation,"cold inflation", has two regime: Slow-roll and (p)reheating.
In the slow-roll limit kinematic energy is small compared to the potential energy term and the universe expands.  Interaction between scalar field (inflaton) and other fields (massive and radiation fields) are neglected. After this period, Kinetic energy is comparable to the potential energy in (p)reheating epoch. In this era inflaton oscillates around the minimum of the potential while losing their energy to other fields (radiation, massless fields) which are presented in the theory. After reheating, the universe is filled by radiation. In (p)reheating epoch, observed universe attaches to the end of inflationary period. Another view of reheating is based on quantum mechanical production of massive particles in classical background inflaton \cite{p1,p2}.
Preheating is probably the most efficient and plausible bridge that could connect inflation to a hot radiation dominated universe \cite{p3,p4}.\\
In warm inflationary scenario 
radiation production occurs during the slow-roll inflation epoch and
(p)reheating is avoided \cite{3}. 
In this scenario  thermal fluctuations could play a
dominant role to produce initial fluctuations which are necessary
for LSS formation \cite{3-i}.
Warm inflationary period ends when the universe stops
inflating. After this period the universe enters in the radiation
phase \cite{3}. Some extensions of this model are found in Ref.\cite{new}.\\
In warm inflation there has to be continuously particle production.
For this to be possible, then the microscopic processes that produce
these particles must occur at a timescale much faster than Hubble
expansion.  Thus the decay rates $\Gamma_i$ (not to be confused with the
dissipative coefficient) must be bigger than $H$. Also these produced
particles must thermalize. Thus the scattering processes amongst these
produced particles must occur at a rate bigger than $H$.
These adiabatic conditions were outlined since the early warm inflation
papers, such as Ref. \cite{1-ne}.  More recently
there has been considerable explicit calculation from Quantum Field Theory (QFT) that
explicitly computes all these relevant decay and scattering rates
in warm inflation models \cite{4nn,arj}.\\
The inflation era in the early evolution of the universe could be described
by tachyonic field, associated with unstable D-brane, because of the tachyon
condensation near the maximum of the effective potential \cite{4-i,5-i,5-i1}. At the late
times, tachyonic fields may add a non-relativistic fluid or a new form of
cosmological dark matter to the universe\cite{1}.
The tachyon inflation is a k-inflation model
\cite{n-1}, for scalar field $\phi$ with a positive potential
$V(\phi)$. Tachyon potentials have two special properties,
firstly a maximum of these potential is obtained where
$\phi\rightarrow 0$ and second property is the minimum of these
potentials is obtained where $\phi\rightarrow \infty$. If the
tachyon field starts to roll down the potential, then universe
dominated by a new form of matter will smoothly evolve from
inflationary universe to an era which is dominated by a
non-relativistic fluid \cite{1}. So, we can explain the phase
of acceleration expansion (inflation) in terms of tachyon field.
Tachyon fields in the ordinary (cold) tachyon inflation framework, after slow-roll epoch, evolves towards  minimum of the potential without oscillating about it \cite{5-i1}, so, the (p)reheating mechanism in cold tachyon inflation does not work.
 Warm tachyon inflation is a picture, where there are dissipative effects play during inflation. As a result of this the inflation evolves in a thermal radiation bath, therefore the reheating problem of cold tachyon inflation \cite{5-i1} can be solved in the framework of warm tachyon inflation.
We note that, the cold tachyonic  inflation era can naturally end with the collision of the two branes. In this situation we do not need warm inflation. If the collision of two branes does not arise naturally, warm inflation is perfectly good scenario that can solve the problem of end of thachyon inflation.\\ 
 
We may live on a brane which is embedded in a higher
dimensional universe. This realization has significant implications to cosmology
\cite{1-f}. In this scenario, which is motivated by string theory, gravity (closed string modes)
can propagates in the bulk, while the standard model of particles (matter fields which are related to open string modes) is confined
to the lower-dimensional brane \cite{2-f}. In term of  Randall-Sundrum suggestion, there are two similar but  phenomenologically different brane world scenarios \cite{rs1,rs2}. In this paper we will consider the brane world model corresponds to the Randall-Sundrum II brane world \cite{rs2}.

 The brane world picture is described by the following action \cite{1-f}
\begin{equation}\label{}
S=\frac{1}{\kappa^2}\int d^5x \sqrt{-g}(R^5+\frac{12}{l^2})-\lambda\int d^4x \sqrt{-g_{brane}}+\int d^4x\sqrt{-g_{brane}}L_{matter}
\end{equation}
In this scenario we have a 3-brane universe which is located in the $5D$ Anti-de Sitter (AdS) spacetime, where this sapacetime is effectively  compactified with curvature scale $l$ of AdS space-time.
$R^5$ is the Ricci scalar in five dimension and $\kappa=8\pi G_5=\frac{8\pi}{M_5^3}$, where $G_5$ is the $5D$ Newton's constant and $M_5$ is Planck scale in five dimensions. $\lambda$ is the tension of the brane and if we have no matter on the brane  $\kappa^2\lambda=\frac{6}{l}$, where the brane becomes Minkowski space-time. In the brane-world model the gravity could propagate in the $5D$ space-time and the Newtonian gravity in four dimensions is reproduced at the scales larger than $l$ on the brane.
$4D$ Einstein's equation projected onto the brane have been found in Ref.\cite{3-f}. Friedmann equation and the equations of linear perturbation theory \cite{4-f} may be modified by these projections.   Einstein's equations which are projected onto the brane with cosmological constant and matter fields which are confined to 3-brane, have the following form \cite{3-f}
\begin{equation}\label{1}
G_{\mu\nu}=-\Lambda_4 g_{\mu\nu}+(\frac{8\pi}{M_4^2})T_{\mu\nu}+(\frac{8\pi}{M_5^3})^2\pi_{\mu\nu}-E_{\mu\nu}
\end{equation}
where $E_{\mu\nu}$ is a projection of 5D weyl tensor, $T_{\mu\nu}$ is energy density tensor on the brane, $M_4=\sqrt{l M_5^3}$ is the Planck scales in 4D and $\pi_{\mu\nu}$ is a tensor quadratic in $T_{\mu\nu}$
\begin{equation}\label{1}
\pi_{\mu\nu}=-\frac{1}{4}T_{\mu\alpha}T^{\alpha}_{\nu}+\frac{1}{12}T_{\alpha}^{\alpha}T_{\mu\nu}+\frac{1}{24}(3T_{\alpha\beta}T^{\alpha\beta}-(T^{\alpha}_{\alpha})^2)g_{\mu\nu}
\end{equation}

Cosmological constant $\Lambda_4$ on the brane in term of 3-brane tension $\lambda$ and $5D$ cosmological constant $\Lambda$ is given by
\begin{equation}\label{}
\nonumber
\Lambda_4=\frac{4\pi}{M_5^3}(\Lambda+\frac{4\pi}{3M_5^3}\lambda^2)
\end{equation}
$4D$ Planck scale is determined by $5D$ Planck scale as
\begin{equation}\label{}
\nonumber
M_4=\sqrt{\frac{3}{4\pi}}(\frac{M_5^2}{\sqrt{\lambda}})M_5=\sqrt{\frac{4\pi\lambda}{3}}l
\end{equation}
The natural boundary conditions to specify the perturbations of this model are imposed, where the perturbations do not diverge at the horizon of the AdS  spacetime and we assume that the Weyl curvature may be neglected.
On the large scale, the behavior of cosmological perturbations on the brane world models is the same as that a closed system on the brane without the effects of the perturbations along the extra-dimensions in the bulk \cite{nn-1}. On the large scale limit, the perturbation parameters of inflation models have a complete set of perturbed equations on the brane which may be solved in quasi-stable and slow-roll limit \cite{nn-1},\cite{6-f}.
The study of the  perturbation evolution  of warm inflation in the brane-world model on the large scale, by using equations solely on the brane and without solving the bulk perturbations is found in Ref.\cite{6-f}. This model has a complete set of perturbed equation on the brane.  We would like to study the warm tachyon inflation model on the brane using this approach. Therefore we will consider the linear cosmological perturbations theory for warm tachyon inflation model on the brane.
In spatially flat FRW model the Friedmann equation, by using Einstein's equation (\ref{1}), has following form \cite{3-f}
\begin{equation}\label{fri}
H^2=\frac{\Lambda_4}{3}+(\frac{8\pi}{3M_4^2})\rho+(\frac{4\pi}{3M_5^3})\rho^2+\frac{\varepsilon}{a^4}
\end{equation}
where $a$ is scale factor of the model and $H$ is Hubble parameter  and $\rho$ is the total energy density on the brane. The last term in the above equation denotes the influence of the bulk gravitons on the brane, where the $\varepsilon$ is an integration constant which arising from Weyl tensor $E_{\mu\nu}$. This term may be rapidly diluted once inflation begins  and we will neglect it. Therefore the projected Weyl tensor term in the effective Einstein equation may be neglected and this term do not give the significant contributions to the observable perturbation parameters. We will also take the $\Lambda_4$ to be vanished at least in the early universe. So, the Friedmann equation reduces  to
\begin{equation}\label{fri1}
H^2=\frac{8\pi}{3M_4^2}\rho(1+\frac{\rho}{2\lambda})
\end{equation}
The brane tension $\lambda$ has been constrained from nucleosynthesis \cite{ten1} $\lambda>(1MeV)^4$ and a stronger limit of it results from current tests for deviation from $Newton^{,} s$ low,$\lambda\geq (10 TeV)^4$, \cite{ten2}.\\  
In the warm inflationary models where, the total energy density $\rho=\rho_{\phi}+\rho_{\gamma}$ is presented on the brane \cite{5-f}, where $\rho_{\gamma}$ is the energy density of the radiation. The  Friedmann equation has this form
\begin{equation}\label{}
\nonumber
H^2=\frac{8\pi}{3M_4^2}(\rho_{\phi}+\rho_{\gamma})(1+\frac{\rho_{\phi}+\rho_{\gamma}}{2\lambda})
\end{equation}
Cosmological perturbations of warm inflation model have been studied in Ref.\cite{9-f}.
Warm tachyon inflationary universe model  has been studied in
Ref.\cite{1-m}, also warm inflation on the brane has been studied in Ref \cite{6-f}. Inflation era is located in a period of dynamical evolution of the universe that the effect of string/M-theory is relevant. On the other hands, string/M-theory is related to higher dimension theories such as space-like branes \cite{4-i}. Therefore in the present work we will study warm-tachyon
inspired inflation in the context of a higher-dimensional
theory instead General Relativity i.e. Randall-Sundrum brane world and cosmological perturbations of the model by using
the above modified Einstein and Friedmann
equations.\\ Recently there has been a new perspective of warm inflation \cite{Clifton:2014fja} which is considered warm inflationary era as a quasi-de Sitter epoch of universe expansion, on the other hands as we mentioned it is believed that we may live on the brane, therefore we interest to study warm tachyon inflation on the brane by using quasi-de Sitter solutions of scale factor. \\           
In one sector of the present work, we would like to consider warm tachyon model on the brane in the context of "intermediate inflation". This scenario is one of the exact solutions of inflationary field equation in the Einstein theory with scale factor $a(t)=a_0\exp(At^f)$ ($A>0, 0<f<1$), this solution of the scale factor in the context of a modified tensor-scalar theory has been found in \cite{Cid:2015pja}. The study of this model is motivated by string/M-theory \cite{4-m}.  If we add the higher order curvature correction, which is proportional to Gauss-Bonnent (GB) term, to Einstein-Hilbert action then we obtain a free-ghost action \cite{5-m}. Gauss-Bonnent interaction is leading order of the "$\alpha$" expansion to low-energy string effective action \cite{5-m} ($\alpha$ is inverse string tension). This theory may be applied for black hole solutions \cite{6-m}, acceleration of the late time universe \cite{7-m} and initial singularity problems \cite{8-m}. The GB interaction in $4D$ with dynamical dilatonic scalar coupling leads to an intermediate form of scale factor \cite{4-m}. Expansion of the universe in the intermediate inflation scenario is slower than standard de sitter inflation with scale factor $a=a_0\exp(H_0 t)$ ($a_0, H_0>0$) which arises as $f=1$, but faster than power-low inflation with scale factor $a=t^p$ ($p>1$). Harrison-Zeldovich \cite{10-m}  spectrum of density perturbation i.e. $n_s=1$ for intermediate inflation models  driven by scalar field is presented for exact values of parameter $f$ \cite{11-m}.\\
On the other hand we will also study our model in the context of "logamediate inflation" with scale factor $a(t)=a_0 \exp(a[\ln t]^{\nu})$ ($\nu >1, A>0$) \cite{12-m}. This model is converted to power-law inflation for $\nu=1$ cases. This scenario is applied in a number of scalar-tensor theories \cite{13-m}. The study of logamediate scenario is motivated by imposing weak general conditions on the cosmological models which have indefinite expansion \cite{12-m}. The effective potential of the logamediate model has been considered in dark energy models \cite{14-m}. This form of potential are also used in supergravity, Kaluza-Klein theories and super-string models \cite{13-m,15-m}. For logamediate models the power spectrum could be either red or blue tilted \cite{16-m}. In Ref.\cite{12-m}, we can find eight possible asymptotic scale factor solutions for cosmological dynamics. Three of these solutions are non-inflationary scale factor, another three one's of solutions give power-low, de sitter and intermediate scale factors. Finally, two cases of these solutions have asymptotic expansion with logamediate scale factor. We will study our model using intermediate and logamediate scenarios.\\
Warm inflation models based on ordinary scalar fields have been studied in \cite{6-f,nik}. Particular model of warm inflation which is driven by tachyon field can be found in \cite{1-m}. In Ref.\cite{Cid:2015ota}, the consistency of warm tachyon inflation with viscous pressure has been studied and the stability analysis for that model has been done. In the present paper we will study warm tachyon inflation without viscosity effect on the brane. We also extended our model by using exact solutions of the scale factor by Barrow \cite{12-m} i.e. inter(loga)mediate solution.\\ 
The paper is organized as: In the next section we will
describe warm-tachyon inflationary universe model in the
brane scenario in the background level. In section (3) we present the perturbation parameters
for our model. In section (4) we study our model  using the
exponential potential in high dissipative regime and high energy limit. In section (5) we study the model using intermediate scenario. In section (6) we develop our model in the context of logamediate inflation.  Finally in
section (7) we close by some concluding  remarks.

\section{The model}
Tachyon scalar field $\phi$ is described by relativistic Lagrangian \cite{1} as:
\begin{equation}\label{}
\nonumber
L=\sqrt{-g}(\frac{R}{16\pi G}-V(\phi)\sqrt{1-g^{\mu\nu}\partial_{\mu}\phi\partial_{\nu}\phi}).
\end{equation}
The stress-energy tensor in a spatially flat Friedmann Robertson Walker
(FRW)
space-time, is presented by
\begin{equation}\label{}
\nonumber
T^{\mu}_{\nu}=\frac{\partial
L}{\partial(\partial_{\mu}\phi)}\partial_{\nu}\phi-g^{\mu}_{\nu}L=diag(-\rho_{\phi},P_{\phi},P_{\phi},P_{\phi}).
\end{equation}
From the above equation, energy density and pressure for a spatially
homogeneous field have the  following
forms:
\begin{equation}\label{rho}
\rho_{\phi}=\frac{V(\phi)}{\sqrt{1-\dot{\phi}^2}}~~~~~~~~~~~~~~P_{\phi}=-V(\phi)\sqrt{1-\dot{\phi}^2},
\end{equation}
where $V(\phi)$ is a  scalar potential
associated with the tachyon field $\phi$. Important characteristics
of this potential are $\frac{dV}{d\phi}<0,$ and $V(\phi\rightarrow
 \infty)\rightarrow 0$  \cite{2}.
In this section, we will present the characteristics of warm tachyon inflation model on the brane in the background level. This model may be described  by an effective fluid where the energy-momentum tensor of this fluid was recognized in the above equation.

The dynamic of the warm tachyon
inflation in spatially flat FRW model on the brane is described by these equations.
\begin{eqnarray}\label{4}
H^2=\frac{8\pi}{3M_4^2}[\frac{V(\phi)}{\sqrt{1-\dot{\phi}^2}}+\rho_{\gamma}][1+\frac{1}{2\lambda}(\frac{V(\phi)}{\sqrt{1-\dot{\phi}^2}}+\rho_{\gamma})]
\end{eqnarray}

\begin{equation}\label{5}
\dot{\rho}_{\phi}+3H(P_{\phi}+\rho_{\phi})=-\Gamma\dot{\phi}^2\Rightarrow
\frac{\ddot{\phi}}{1-\dot{\phi}^2}+3H\dot{\phi}+\frac{V'}{V}=-\frac{\Gamma}{V}\sqrt{1-\dot{\phi}^2}\dot{\phi}
\end{equation}

and
\begin{equation}\label{6}
\dot{\rho}_{\gamma}+4H\rho_{\gamma}=\Gamma\dot{\phi}^2
\end{equation}

where $\Gamma$ is the dissipative coefficient. In the above equations dots "." mean derivative with
respect to cosmic time and prime  denotes derivative with respect
to scalar field $\phi$. During slow-roll inflation era the energy density
(\ref{rho}) is the order of potential $\rho_{\phi}\sim V$ and
dominates over the radiation energy $\rho_{\phi}>\rho_{\gamma}$.
Using the slow-roll limit when $\dot{\phi}\ll 1$ and
$\ddot{\phi}\ll(3H+\frac{\Gamma}{V})\dot{\phi}$ \cite{3}, and also when the inflation
radiation production is quasi-stable ($\dot{\rho}_{\gamma}\ll
4H\Gamma$, $\dot{\rho}_{\gamma}\ll\Gamma\dot{\phi}^2$), the
dynamic equations (\ref{4}) and (\ref{5})   are reduced to
\begin{equation}\label{7}
H^2=\frac{8\pi}{3M_4^2}V(1+\frac{V}{2\lambda})
\end{equation}

\begin{equation}\label{8}
3H(1+r)\dot{\phi}=-\frac{V'}{V}
\end{equation}

where $r=\frac{\Gamma}{3HV}$. In canonical  warm inflation scenario the relative strength of thermal damping ($\Gamma$) should be compared to expansion damping ($H$). We must analysis the warm inflation model in background and linear perturbation levels on our expanding over timescales which are shorter than the variation of expansion rate, but large compared to the microphysical processes 
\begin{equation}
\nonumber
\frac{V}{\Gamma}\ll\tau\ll H^{-1}, \Rightarrow \Gamma\gg HV 
\end{equation}
For more discussion plz see the apendix.
Particle production in fact tacks place at a constant rate during warm inflation for canonical scalar field where strength of thermal damping dominates over the
effect of expansion damping ($\Gamma>H$) but for tachyonin scalar fields as be presented in the above equation $\Gamma> HV$. We will study our model in high  dissipative regime ($r\gg 1$). Using these conditions we have $\Gamma\gg HV$ which agrees with particle production condition ($\Gamma>VH$).

 From 
Eqs.(\ref{6}),(\ref{7}), and (\ref{8}), $\rho_{\gamma}$ could be written as
\begin{equation}\label{9}
\rho_{\gamma}=\frac{\Gamma\dot{\phi}^2}{4H}=\frac{M_4^2r}{32\pi(1+r)^2(1+\frac{V}{2\lambda})}(\frac{V'}{V})^2=\sigma
T_r^4
\end{equation}

where  $T_r$ is the temperature of thermal bath and $\sigma$ is
Stefan-Boltzmann constant. We introduce the slow-roll parameters
for our model as
\begin{eqnarray}\label{10}
\epsilon=-\frac{\dot{H}}{H^2}\simeq\frac{M_4^2}{16\pi}\frac{V'^2}{(1+r)V^3}\frac{1+\frac{V}{\lambda}}{(1+\frac{V}{2\lambda})^2}
\end{eqnarray}

and
\begin{eqnarray}\label{11}
\eta=-\frac{\ddot{H}}{H\dot{H}}\simeq\frac{M_4^2}{8\pi}\frac{V'}{V^2(1+r)[1+\frac{V}{2\lambda}]}~~~~~~~~~~~~~~~~~~~~\\
\nonumber \times
[\frac{2V''}{V'}-\frac{V'}{V}-\frac{r'}{(1+r)}+\frac{V'}{\lambda+V}]-2\epsilon
\end{eqnarray}

A relation between two energy densities $\rho_{\phi}$ and
$\rho_{\gamma}$ is obtained from Eqs. (\ref{9}) and (\ref{10})
\begin{equation}\label{12}
\rho_{\gamma}=\frac{r}{2(1+r)}\frac{[1+
\frac{\rho_{\phi}}{2\lambda}]}{[1+\frac{\rho_{\phi}}{\lambda}]}\rho_{\phi}\epsilon\simeq\frac{r}{2(1+r)}
\frac{[1+\frac{V}{2\lambda}]}{[1+\frac{V}{\lambda}]}V\epsilon
\end{equation}

The condition of inflation epoch $\ddot{a}>0$ could be obtained
by inequality $\epsilon<1$. Therefore from above equation,
warm-tachyon inflation on the brane
could take place when
\begin{equation}\label{13}
\frac{2(1+r)}{r}\rho_{\gamma}<\frac{1+\frac{\rho_{\phi}}{2\lambda}}{1+\frac{\rho_{\phi}}{\lambda}}\rho_{\phi}
\end{equation}

Inflation period ends when $\epsilon\simeq 1$ which implies
\begin{equation}\label{14}
\frac{M_4^2}{8\pi}[\frac{V'_f}{V_f}]^2\frac{1+\frac{V_f}{\lambda}}{(1+\frac{V_f}{2\lambda})^2}\frac{1}{V_f}\simeq
2(1+r_f)
\end{equation}

where the subscript $f$ denotes the end of inflation. The number
of e-folds is given by
\begin{eqnarray}\label{15}
N=\int_{\phi_{*}}^{\phi_f}Hdt=\int_{\phi_{*}}^{\phi_f}\frac{H}{\dot{\phi}}d\phi=-\frac{8\pi}{M_4^2}\int_{\phi_{*}}^{\phi_f}\frac{V^2}{V'}(1+r)[1+\frac{V}{2\lambda}]d\phi
\end{eqnarray}

where the subscript $*$ denotes the epoch when the cosmological
scale exits the horizon.
\section{Perturbation}
In this section we will study inhomogeneous perturbations of the FRW background. As we have mentioned in the introduction we ignore the influence of the bulk gravitons on the brane which arising from Weyl tensor $E_{\mu\nu}$, so we neglect the back-reaction due to metric perturbations in the fifth dimension.
 These perturbations in the longitudinal gauge, may be described by the perturbed FRW metric
\begin{equation}\label{16}
ds^2=(1+2\Phi)dt^2-a^2(t)(1-2\Psi)\delta_{ij}dx^idx^j
\end{equation}

where $\Phi$ and $\Psi$ are gauge-invariant metric perturbation variables \cite{7-f}.
The equation of motion is given by

 \begin{eqnarray}\label{}
\frac{\ddot{\delta\phi}}{1-\dot{\phi}^2}+[3H+\frac{\Gamma}{V}]\dot{\delta\phi}+[-a^{-2}\nabla^2+(\frac{V'}{V})'+\dot{\phi}(\frac{\Gamma}{V})']\delta\phi\\
\nonumber
-[\frac{1}{1-\dot{\phi}^2}+3]\dot{\phi}\dot{\Phi}-[\dot{\phi}\frac{\Gamma}{V}-2 \frac{V'}{V}]\Phi=0
\end{eqnarray}
We expand the small change of field $\delta\phi$ into Fourier components as
\begin{eqnarray}\label{}
\delta\phi(x)=\int \frac{d^3k}{(2\pi)^3}[e^{ikx}\delta\phi(k,t)a_k+e^{-ikx}\delta\phi(k,t)a_k^{*}]
\end{eqnarray}
In warm inflation
thermal fluctuations of the inflation dominate over the quantum ones, therefore we have classical perturbation of scalar field $\delta \phi$. 
All perturbed quantities have a spatial sector of the form $e^{i\mathbf{kx}}$, where $k$ is the wave number. Perturbed Einstein field equations in momentum space have only the temporal parts
\begin{equation}\label{}
\nonumber
\Phi=\Psi
\end{equation}

\begin{equation}\label{17}
\dot{\Phi}+H\Phi=\frac{4\pi}{M_4^2}[-\frac{4\rho_{\gamma}av}{3k}+\frac{V\dot{\phi}}{\sqrt{1-\dot{\phi}^2}}\delta\phi][1+\frac{1}{\lambda}[\rho_{\gamma}+\frac{V}{\sqrt{1-\dot{\phi}^2}}]]
\end{equation}

\begin{eqnarray}\label{18}
\frac{\ddot{\delta\phi}}{1-\dot{\phi}^2}+[3H+\frac{\Gamma}{V}]\dot{\delta\phi}+[\frac{k^2}{a^2}+(\frac{V'}{V})'+\dot{\phi}(\frac{\Gamma}{V})']\delta\phi\\
\nonumber
-[\frac{1}{1-\dot{\phi}^2}+3]\dot{\phi}\dot{\Phi}-[\dot{\phi}\frac{\Gamma}{V}-2\frac{V'}{V}]\Phi=0
\end{eqnarray}

\begin{eqnarray}\label{19}
(\dot{\delta\rho_{\gamma}})+4H\delta\rho_{\gamma}+\frac{4}{3}ka\rho_{\gamma}v-4\rho_{\gamma}\dot{\Phi}-\dot{\phi}^2\Gamma'\delta\phi-\Gamma\dot{\phi}^2[2(\dot{\delta\phi})-3\dot{\phi}\Phi]=0
\end{eqnarray}

and
\begin{equation}\label{20}
\dot{v}+4Hv+\frac{k}{a}[\Phi+\frac{\delta\rho_{\gamma}}{4\rho_{\gamma}}+\frac{3\Gamma\dot{\phi}}{4\rho_{\gamma}}\delta\phi]=0
\end{equation}

The above equations are obtained for Fourier components $e^{i\mathbf{kx}}$, where the subscript $k$ is omitted. $v$ in the above set of equations is presented by  the decomposition of the velocity field ($\delta u_j=-\frac{iak_J}{k}ve^{i\mathbf{kx}}, j=1,2,3$) \cite{7-f}.

Note that the effect of the bulk (extra dimension) to perturbed projected Einstein field equations on the brane may be found in Eq.(\ref{17}).
We will describe the non-decreasing adiabatic and isocurvature modes of our model on large scale limit. In this limit we have obtained a complete set of perturbation equations on the brane. Therefore the perturbation variables along the extra-dimensions in the bulk could not have any contribution to the perturbation  equations on super-horizon scales (see for example \cite{nn-1},\cite{6-f}.). The same approach, for non-tachyon warm inflation model on the brane,  in Ref.\cite{6-f} is presented.
Warm inflation model may be considered as a hybrid-like inflationary model where the inflaton field interacts with radiation field \cite{9-f}, \cite{8-f}. Entropy perturbation may be related to dissipation term \cite{10-f}.
Perturbation of entropy in warm inflation model is given by \cite{nn-3}
\begin{equation}\label{}
\delta S=e=-V_{,\phi T}\delta\phi-V_{,TT}\delta T
\end{equation}
In this paper we will study potential of the model as a function of scalar field ($V(\phi)$), therefore the entropy perturbation will be neglected. We will study this important issue (potential as function of temperature, $V(\phi,T)$) in future works.

 During inflationary phase with slow-roll approximation, for non-decreasing adiabatic modes on large scale limit $k\ll aH$, we assume that the perturbed quantities could not vary strongly. So we have $H\Phi\gg\dot{\Phi}$, $(\ddot{\delta\phi})\ll(\Gamma+3H)(\dot{\delta\phi})$, $(\dot{\delta\rho_{\gamma}})\ll\delta\rho_{\gamma}$ and $\dot{v}\ll 4Hv$. In the slow-roll limit and by using the above limitations, the set of perturbed equations are reduced to
\begin{equation}\label{21}
\Phi\simeq\frac{4\pi}{HM_4^2}[-\frac{4\rho_{\gamma}av}{3k}+V\dot{\phi}\delta\phi][1+\frac{V}{\lambda}]
\end{equation}

\begin{equation}\label{22}
[3H+\frac{\Gamma}{V}]\dot{\delta\phi}+[(\frac{V'}{V})'+\dot{\phi}(\frac{\Gamma}{V})']\delta\phi
\simeq[\dot{\phi}\frac{\Gamma}{V}-2(\frac{V'}{V})]\Phi
\end{equation}

\begin{equation}\label{23}
\frac{\delta\rho_{\gamma}}{\rho_{\gamma}}\simeq\frac{\Gamma'}{\Gamma}\delta\phi-3\Phi
\end{equation}
and
\begin{eqnarray}\label{24}
v\simeq-\frac{k}{4aH}(\Phi+\frac{\delta\rho_{\gamma}}{4\rho_{\gamma}}+\frac{3\Gamma\dot{\phi}}{4\rho_{\gamma}}\delta\phi)
\end{eqnarray}
Using Eqs.(\ref{21}), (\ref{23}) and (\ref{24}), perturbation variable $\Phi$ is determined
\begin{eqnarray}\label{25}
\Phi=\frac{4\pi}{M_4^2}(\frac{V\dot{\phi}}{H})[1+\frac{\Gamma}{4HV}+\frac{\Gamma'\dot{\phi}}{48H^2V}](1+\frac{V}{\lambda})\delta\phi
\end{eqnarray}

We can solve the above equations by taking tachyon field $\phi$ as the independent variable in place of cosmic time $t$. Using Eq.(\ref{8}) we find
\begin{eqnarray}\label{26}
(3H+\frac{\Gamma}{V})\frac{d}{dt}=(3H+\frac{\Gamma}{V})\dot{\phi}\frac{d}{d\phi}=-\frac{V'}{V}\frac{d}{d\phi}
\end{eqnarray}

From above equation, Eq.(\ref{22}) and Eq.(\ref{25}), the expression $\frac{(\delta\phi)'}{\delta\phi}$ is obtained
\begin{eqnarray}\label{27}
\frac{(\delta\phi)'}{\delta\phi}=\frac{V}{ V'}[(\frac{V'}{V})'+\dot{\phi}(\frac{\Gamma}{V})'+\frac{4\pi}{M_4^2}(-\dot{\phi}\frac{\Gamma}{V}+2(\frac{V'}{V})')\\
\nonumber
\times(\frac{V\dot{\phi}}{H})[1+\frac{\Gamma}{4HV}+\frac{\Gamma'\dot{\phi}}{48H^2V}](1+\frac{V}{\lambda})]~~~~~~~~~~~~~~
\end{eqnarray}

We will return to the above relation. Following Refs.\cite{1-m}, \cite{6-f},  \cite{10-f},  we introduce auxiliary function $\chi$ as
\begin{equation}\label{28}
\chi=\frac{V\delta\phi}{V'}\exp[\int\frac{1}{3H+\frac{\Gamma}{V}}(\frac{\Gamma}{V})'d\phi]
\end{equation}

From above definition we have
\begin{eqnarray}\label{29}
\frac{\chi'}{\chi}=\frac{(\delta\phi)'}{\delta\phi}-\frac{V}{V'}(\frac{V'}{V})'+\frac{(\frac{\Gamma}{V})'}{3H+\frac{\Gamma}{V}}
\end{eqnarray}

Using above equation and Eq.(\ref{27}), we find
\begin{equation}\label{30}
\frac{\chi'}{\chi}=\frac{4\pi}{M_4^2}(-\frac{V\dot{\phi}}{ V'}\frac{\Gamma}{V}+2)
(\frac{V\dot{\phi}}{H})[1+\frac{\Gamma}{4HV}+\frac{\Gamma'\dot{\phi}}{48H^2V}](1+\frac{V}{\lambda})
\end{equation}

We could rewrite this equation, using Eqs.(\ref{7}) and (\ref{8})
\begin{equation}\label{31}
\frac{\chi'}{\chi}=-\frac{9}{8}\frac{2H+\frac{\Gamma}{V}}{(3H+\frac{\Gamma}{V})^2}(\Gamma+4HV-\frac{\Gamma' V'/V}{12H(3H+\frac{\Gamma}{V})})\frac{ V'}{V^2}\frac{[1+\frac{V}{\lambda}]}{1+\frac{V}{2\lambda}}
\end{equation}

A solution for the above equation is
\begin{eqnarray}\label{32}
\chi(\phi)=C\exp(-\int\{-\frac{9}{8}\frac{2H+\frac{\Gamma}{V}}{(3H+\frac{\Gamma}{V})^2}~~~~~~~~~~~~~~~~~~\\
\nonumber \times(\Gamma+4HV-\frac{\Gamma' V'/V}{12H(3H+\frac{\Gamma}{V})})\frac{ V'}{V^2}\frac{[1+\frac{V}{\lambda}]}{1+\frac{V}{2\lambda}}\}d\phi)
\end{eqnarray}

where $C$ is integration constant. From above equation and Eq.(\ref{29}) we find small change of variable $\delta\phi$ as:
\begin{equation}\label{33}
\delta\phi=C\frac{V'}{V} \exp(\Im(\phi))
\end{equation}

where
\begin{eqnarray}\label{34}
\Im(\phi)=-\int[\frac{(\frac{\Gamma}{V})'}{3H+\frac{\Gamma}{V}}+(\frac{9}{8}\frac{2H+\frac{\Gamma}{V}}{(3H+\frac{\Gamma}{V})^2}~~~~~~~~~~~~~~~~~~~~~~\\
\nonumber
\times(\Gamma+4HV-\frac{\Gamma' V'/V}{12H(3H+\frac{\Gamma}{V})})\frac{ V'}{V^2}\frac{[1+\frac{V}{\lambda}]}{1+\frac{V}{2\lambda}})]d\phi
\end{eqnarray}

In the above calculations we have used the perturbation methods in warm inflation models \cite{1-m}, \cite{6-f}, \cite{10-f}, where the small change of variable $\delta\phi$ may be generated by thermal fluctuations instead of quantum fluctuations \cite{5},   and the integration constant $C$ may be driven by boundary conditions for field perturbation. Perturbed matter fields of our model are inflaton $\delta\phi$, radiation $\delta\rho_r$ and velocity $k^{-1}(P+\rho)v_{,i}$. We can explain the cosmological perturbations in terms of gauge-invariant variables. These variables are important for development of perturbation after the end of inflation period. The curvature perturbation $\mathfrak{R}$ and entropy perturbation $e$ are defied by \cite{nn-2}
\begin{eqnarray}\label{}
\mathfrak{R}=\Phi-k^{-1}aHv~~~~~~~\\
\nonumber
e=\delta P-c_s^2\delta\rho~~~~~~
\end{eqnarray}
where $c_s^2=\frac{\dot{P}}{\dot{\rho}}$. The boundary condition of warm inflation models are found in very large scale limits i.e., $k\ll aH$ where the curvature perturbation $\mathfrak{R}\sim const$ and the entropy perturbation vanishes \cite{nn-3}.

Finally the density perturbation is given by \cite{12-f}
\begin{equation}\label{35}
\delta_H=\frac{2}{5}M_4^2\frac{V\exp(-\Im(\phi))}{ V'}\delta\phi=\frac{2}{15}M_4^2\frac{\exp(-\Im(\phi))}{Hr\dot{\phi}}\delta\phi
\end{equation}

For high or low energy limit ($V\gg\lambda$ or $V\ll \lambda$) and by inserting $\Gamma=0$, the above equation reduces to $\delta_{H}\simeq\frac{H}{\dot{\phi}}\delta\phi$ which agrees with the density perturbation in cold inflation model \cite{1-i}. In the warm inflation model the fluctuations of the scalar field in high dissipative regime ($r\gg 1$) may be generated by thermal fluctuation instead of quantum fluctuations \cite{5} as:
\begin{equation}\label{36}
(\delta\phi)^2\simeq\frac{k_F T_r}{2\pi^2}
\end{equation}
where in this limit freeze-out wave number $k_F=\sqrt{\frac{\Gamma H}{V}}=H\sqrt{3r}\geq H$ corresponds to the freeze-out scale at the point when, dissipation damps out to thermally excited fluctuations ($\frac{V''}{V'}<\frac{\Gamma H}{V}$) \cite{5}. $\delta\phi$ in Eq.(\ref{36}) can be found in Ref.\cite{5},
where Fourier transformed to momentum space is used (see for example Appendix of Ref.\cite{5} and Sec. 4 of Ref.\cite{11-m}), therefore $\delta\phi$ is introduced in Fourier space and we can present spectral index and running in Fourier space. 
With the help of  Eqs.(\ref{35}) and (\ref{36}) in high energy ($V\gg\lambda$) and high dissipative regime ($r\gg 1$) we find
\begin{equation}\label{37}
\delta_H^2=\frac{2\sqrt{3}}{75\pi^2}M_4^4\frac{\exp(-2\tilde{\Im}(\phi))}{\sqrt{r}\tilde{\epsilon}}\frac{T_r}{H}
\end{equation}

or equivalently
\begin{eqnarray}\label{}
\nonumber
\delta_H^2=\frac{4M_4^5\lambda^{\frac{1}{2}}}{25(2\pi)^{\frac{5}{2}}\sigma^{\frac{1}{4}}}V^{-\frac{3}{4}}r^{-\frac{1}{2}}\epsilon^{-\frac{3}{4}}\exp(-2\tilde{\Im}(\phi))
\end{eqnarray}

where
\begin{equation}\label{38}
\tilde{\Im}(\phi)=-\int[\frac{1}{3Hr}(\frac{\Gamma}{V})'+\frac{9}{4}(1-\frac{(\ln\Gamma)' V'/V}{36rH^2})\frac{V'}{V}]d\phi
\end{equation}

and
\begin{equation}\label{39}
\tilde{\epsilon}=\frac{M_4^2\lambda}{4\pi r}\frac{V'^2}{V^4}
\end{equation}

An important perturbation parameter of inflation models is scalar index $n_s$ which
in high dissipative regime is presented by
\begin{equation}\label{40}
n_s=1+\frac{d\ln \delta_H^2}{d\ln k}\approx
1-\frac{3}{4}\tilde{\epsilon}+\frac{3}{4}\tilde{\eta}+\tilde{\epsilon}(\frac{V}{V'})(2\tilde{\Im}'(\phi)+\frac{r'}{2r})
\end{equation}

where
\begin{equation}\label{41}
\tilde{\eta}=\frac{M_4^2\lambda}{4\pi r}\frac{V'}{V^3}[\frac{2V''}{V'}-\frac{r'}{r}]-2\tilde{\epsilon}
\end{equation}

In Eq.(\ref{40}) we have used a relation between small change of
the number of e-folds and interval in wave number ($dN=-d\ln k$).
Running of the scalar spectral index may be found as
\begin{eqnarray}\label{42}
\alpha_s=\frac{dn_s}{d\ln k}=-\frac{dn_s}{dN}=-\frac{d\phi}{dN}\frac{dn_s}{d\phi}=\frac{M_4^2\lambda}{4\pi r}\frac{V'n_s'}{V^3}
\end{eqnarray}

This parameter is one of the interesting cosmological
perturbation parameters which is approximately $-0.038$, by using
 observational results \cite{6}. During inflation epoch,
there are two independent components of gravitational waves
($h_{\times +}$) with action of massless scalar field are
produced by the generation of tensor perturbations. Tensor perturbations do not couple to the thermal background, therefore gravitational waves are only generated by quantum fluctuations, same as in standard fluctuations \cite{5}. However, if the gravitational sector is modified then the expression for tensor power spectrum changes with respect to General Relativity. In particular, the amplitude
of the tensor perturbation  on the brane is presented as \cite{Langlois:2000ns, Herrera:2015aja}
\begin{eqnarray}\label{43}
A_g^2=\frac{16\pi}{M_4^4}(\frac{H}{2\pi})^2F^2(x)=\frac{16}{3M_4^2\lambda}V^2F^2(x)
\end{eqnarray}

where, the temperature $T$ in extra factor $\coth[\frac{k}{2T}]$,
denotes the temperature of the thermal background of
gravitational wave \cite{7}, $x=[\frac{3H^2M_4^2}{4\pi\lambda}]^{\frac{1}{2}}$ and $F(x)=\lbrace\sqrt{1+x^2}-x^2\sinh^{-1}(\frac{1}{x})\rbrace^{-\frac{1}{2}}$ (In high energy limit, $V\gg\lambda$, we have $F(x)=[\frac{27 M_4^2}{16\pi\lambda}]^{\frac{1}{4}}H^{\frac{1}{2}}=[\frac{3\pi}{\lambda^3 M_4^2}]^{\frac{1}{4}}V^{\frac{1}{2}}$). Spectral index $n_g$ is presented as:
\begin{eqnarray}\label{44}
n_g=\frac{d}{d\ln k}(\ln [\frac{A_g^2}{\coth(\frac{k}{2T})}])\simeq-2\tilde{\epsilon}
\end{eqnarray}
where $A_g\propto k^{n_g}\coth[\frac{k}{2T}]$ \cite{7}.  Using Eqs. (\ref{37})  and
(\ref{43}) we write the tensor-scalar ratio in high dissipative
regime
\begin{eqnarray}\label{45}
R(k)=\frac{A_g^2}{P_R}|_{k=k_{0}}=\frac{16.2^{\frac{5}{2}}\pi^{\frac{11}{4}}\sigma^{\frac{1}{4}} V^{\frac{13}{4}}r^{\frac{1}{2}}\epsilon^{\frac{3}{4}}}{3^{\frac{3}{4}}.M_4^{\frac{15}{2}}\lambda^{\frac{9}{4}}}\exp(2\tilde{\Im}(\phi))\coth(\frac{k}{2T})
\end{eqnarray}

where $k_{0}$ is referred  to pivot point \cite{7} and $P_R=\frac{25}{4}\delta_H^2$. An upper bound for this parameter is given
by using  Planck data, $R<0.11$ \cite{6}.

\section{Exponential potential }
In this section we consider our model with the tachyonic
effective potential
\begin{equation}\label{46}
V(\phi)=V_0\exp(-\alpha\phi)
\end{equation}

where parameter $\alpha>0$  is related to mass
of tachyon field \cite{8}. The exponential form of the potential has
characteristics of tachyon field ($\frac{dV}{d\phi}<0,$ and
$V(\phi\rightarrow 0)\rightarrow V_{max}$ ). We develop our model
in high dissipative regime i.e. $r\gg 1$  and high energy limit  i.e. $V\gg \lambda$ for a constant
dissipation coefficient $\Gamma$. From Eq.(\ref{39}) slow-roll parameter $\tilde{\epsilon}$ in the present case has the form
\begin{equation}\label{47}
\tilde{\epsilon}=\frac{M_4^2\lambda}{8\pi}\frac{\alpha^2}{rV_0^2 e^{-2\alpha\phi}}
\end{equation}

Also the other slow-roll parameter $\tilde{\eta}$ is obtained from Eq.(\ref{41})
\begin{equation}\label{48}
\tilde{\eta}=-\frac{M_4^2}{4\pi}\frac{\alpha^2}{rV_0^2 e^{-2\alpha\phi}}
\end{equation}

Dissipation parameter $r=\frac{\Gamma}{3HV}$ in this case is given by
\begin{equation}\label{49}
r=\sqrt{\frac{\Gamma_0^2M_4^2\lambda}{12\pi}}\frac{e^{2\alpha\phi}}{V_0^2}
\end{equation}

We find the evolution of tachyon field with the help of Eq.(\ref{8})
\begin{equation}\label{50}
\phi(t)=\frac{1}{\alpha}\ln[\frac{\alpha^2 V_0}{\Gamma_0}t+e^{\alpha\phi_i}]
\end{equation}

where $\phi_i=\phi(t=o)$.
Hubble parameter for our model has this form
\begin{equation}\label{51}
H=\sqrt{\frac{4\pi}{3M_4^2\lambda}}V_0e^{-\alpha\phi}
\end{equation}



Using Eqs.(\ref{12}) and (\ref{47}), the energy density of the radiation field in high dissipative limit becomes
\begin{equation}\label{54}
\rho_{\gamma}=\frac{3M_4\alpha^2}{16\Gamma_0}\frac{V_0^2}{\sqrt{3\pi\lambda}}e^{-2\alpha\phi}
\end{equation}

and, in terms of tachyon field energy density $\rho_{\phi}$ becomes
\begin{equation}\label{55}
\rho_{\gamma}=\frac{3M_4^2}{16\sqrt{3\pi\lambda}}(\frac{\alpha^2}{\Gamma_0})\rho_{\phi}^2
\end{equation}

From Eq.(\ref{15}) the number of e-folds, at the end of inflation, by using the potential (\ref{46}) for our inflation model is presented by
\begin{equation}\label{56}
N_{total}=\sqrt{\frac{4\pi\lambda}{3M_4^2}}\frac{\Gamma_0}{\alpha}(\phi_f-\phi_i)
\end{equation}

or equivalently
\begin{equation}\label{57}
N_{total}=\sqrt{\frac{4\pi\lambda}{3M_4^2}}\frac{\Gamma_0}{\alpha^2}\ln(\frac{V_i}{V_f})
\end{equation}

where $V_i>V_f$. Using Eqs.(\ref{37}) and (\ref{45}), we could find the scalar spectrum and scalar-tensor ratio
\begin{equation}\label{58}
\delta_H^2=A\exp(-\frac{7}{2}\alpha\phi)
\end{equation}
where $A=\frac{16\sqrt{3}}{75\pi}\frac{V_0^{\frac{3}{2}}M_4^2}{\alpha^2}(\frac{\Gamma_0^2M_4^2\lambda}{12\pi})^{\frac{1}{2}}(\frac{3M_4^2\lambda}{4\pi})^{\frac{1}{4}}$.
and
\begin{equation}\label{59}
R=B\exp(-\alpha\phi)
\end{equation}
where $B=\frac{50\pi^{\frac{3}{2}}\alpha^2V_0}{3M_4^7\lambda^{\frac{3}{2}}T_r}(\frac{36\pi^2}{\Gamma_0^2M_4^4\lambda^4})^{\frac{1}{4}}$.
In the above equation we have used the Eq.(\ref{38}) where
\begin{equation}\label{60}
\tilde{\Im(\phi)}=-\frac{5}{4}\ln V
\end{equation}

These parameters may by restricted by Planck observational data \cite{6,planck}.

\section{Intermediate inflation}
Intermediate inflation is denoted by the scale factor
\begin{eqnarray}\label{1-i}
a(t)=a_0\exp(At^f),~~~0<f<1
\end{eqnarray}
This model of inflation is faster than power-low inflation and slower than de sitter inflation. In this section we will study our model in the context of intermediate inflation in two cases: 1- $\Gamma=\Gamma_0$ case, 2-$\Gamma=\Gamma_1 V(\phi)$ which have been considered in the literature \cite{1-m}.
\subsection{$\Gamma=\Gamma_0$ case}
In high dissipative ($r\gg1$) and high energy ($V\gg\lambda$) limits the equations of the slow-roll inflation i.e. Eq.(\ref{4}) and (\ref{5}) are simplified as:
\begin{eqnarray}\label{2-i}
V=(\frac{3\lambda M_4^2}{4\pi})^{\frac{1}{2}}H\\
\nonumber
\dot{\phi}^2=-\frac{\dot{V}}{\Gamma}~~~~~~~~
\end{eqnarray}
Inflaton field  may be derived from above equations in this case ($\Gamma=\Gamma_0$)
\begin{eqnarray}\label{}
\phi-\phi_0=\beta t^{\frac{f}{2}}
\end{eqnarray}
where $\beta=(\frac{12\lambda M_4^2 A^2(1-f)^2}{\pi f^2\Gamma_0^2})$. Using above equation and the scale factor of intermediate inflation, tachyonic potential and Hubble parameter  are presented as:
\begin{eqnarray}\label{}
H(\phi)=fA(\frac{\phi-\phi_0}{\beta})^{\frac{2f-2}{f}}~~~~~~~~\\
\nonumber
V(\phi)=(\frac{3\lambda M_4^2 f^2A^2}{4\pi})^{\frac{1}{2}}(\frac{\phi-\phi_0}{\beta})^{\frac{2f-2}{f}}
\end{eqnarray}
Dissipative parameter $r$ is given by using above equation
\begin{eqnarray}\label{}
r=\frac{\Gamma_0}{3HV}=\frac{4\pi \Gamma_0}{9(fA)^2M_4^2\lambda}(\frac{\phi-\phi_0}{\beta})^{\frac{4-4f}{f}}
\end{eqnarray}
The slow-roll parameters of the model in the present case may be obtained as:
\begin{eqnarray}\label{}
\epsilon=-\frac{\dot{H}}{H^2}=\frac{1-f}{fA}(\frac{\phi-\phi_0}{\beta})^{-2}~~~~~~\\
\nonumber
\eta=-\frac{\ddot{H}}{\dot{H}H}=\frac{2-f}{fA}(\frac{\phi-\phi_0}{\beta})^{-2}~~~~~
\end{eqnarray}
We present the number of e-folds as
\begin{eqnarray}\label{}
N=\int_{t_1}^{t}H dt=A([\frac{\phi-\phi_0}{\beta}]^2-[\frac{\phi_1-\phi_0}{\beta}]^2)
\end{eqnarray}
where $\phi_1=\phi_0+\beta(\frac{1-f}{fA})^{\frac{1}{2}}$, is the scalar field at the begining of the inflation. From the above equation we can present the scalar field in term of number of e-folds and intermediate parameters
\begin{eqnarray}\label{}
\phi=\beta(\frac{N}{A}+\frac{1-f}{fA})^{\frac{1}{2}}+\phi_0
\end{eqnarray}
Now we could find the perturbation parameters of the model. The power-spectrum is obtained from Eqs.(\ref{37}), (\ref{38}) and (\ref{60})
\begin{eqnarray}\label{}
P_R=\frac{25}{4}\delta_H^2=\frac{M_4^5\lambda^{\frac{1}{2}}}{(2\pi)^{\frac{5}{2}}\sigma^{\frac{1}{4}}}\frac{V^{\frac{7}{4}}}{r^{\frac{1}{2}}\epsilon^{\frac{3}{4}}}
=A_1(\frac{\phi-\phi_0}{\beta})^{\frac{14f-11}{2f}}\\
\nonumber
=A_1(\frac{N}{A}+\frac{1-f}{fA})^{\frac{14f-11}{4f}}~~~~~~~~~~~~~~~~~~~~~~~
\end{eqnarray}
where $A_1=\frac{2^{\frac{5}{2}}M_4^{\frac{31}{4}}(3\lambda)^{\frac{15}{8}}(fA)^{\frac{7}{2}}}{(4\pi)^{\frac{31}{8}}\sigma^{\frac{1}{4}}\Gamma_0^{\frac{1}{2}}(1-f)^{\frac{3}{4}}}$.
We present the spectral index $n_s$  which is one of the important perturbation parameters from Eqs.(\ref{40}) and (\ref{60})
\begin{eqnarray}\label{}
n_s=1+\frac{3}{4}\eta-\frac{17}{4}\epsilon=1-\frac{11-14f}{4fA}(\frac{\phi-\phi_0}{\beta})^{-2}=1-\frac{11-14f}{4fA}(\frac{N}{A}+\frac{1-f}{fA})^{-1}
\end{eqnarray}
Harrison-Zeldovich spectrum, i.e. $n_s=1,$ is obtained for an exact value of parameter $f$ (i.e. $f=\frac{11}{14}$).
For $f<\frac{11}{14}$ we found the $n_s<1$ cases which is compatible with observational data.

\begin{figure}[h]
\centering
  \includegraphics[width=10cm]{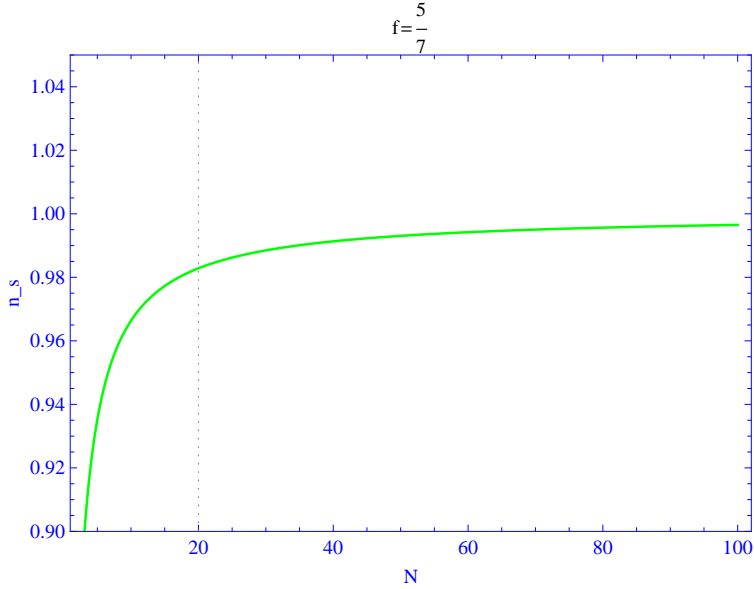}
  \caption{In this graph we plot the spectral index $n_s$ in term of the number of e-folds $N$. We can find best range of spectral index ratio where $N>60$.}
 \label{fig:n-N-int-const}
\end{figure}

In Fig.(\ref{fig:n-N-int-const}),
we plot the spectral index in term of number of e-fold where $f=\frac{5}{7}$. For $N>60$ we can see the spectral index is confined to $0.98<n_s<1$ which is compatible with Planck
data \cite{6}.

\begin{figure}[h]
\centering
  \includegraphics[width=10cm]{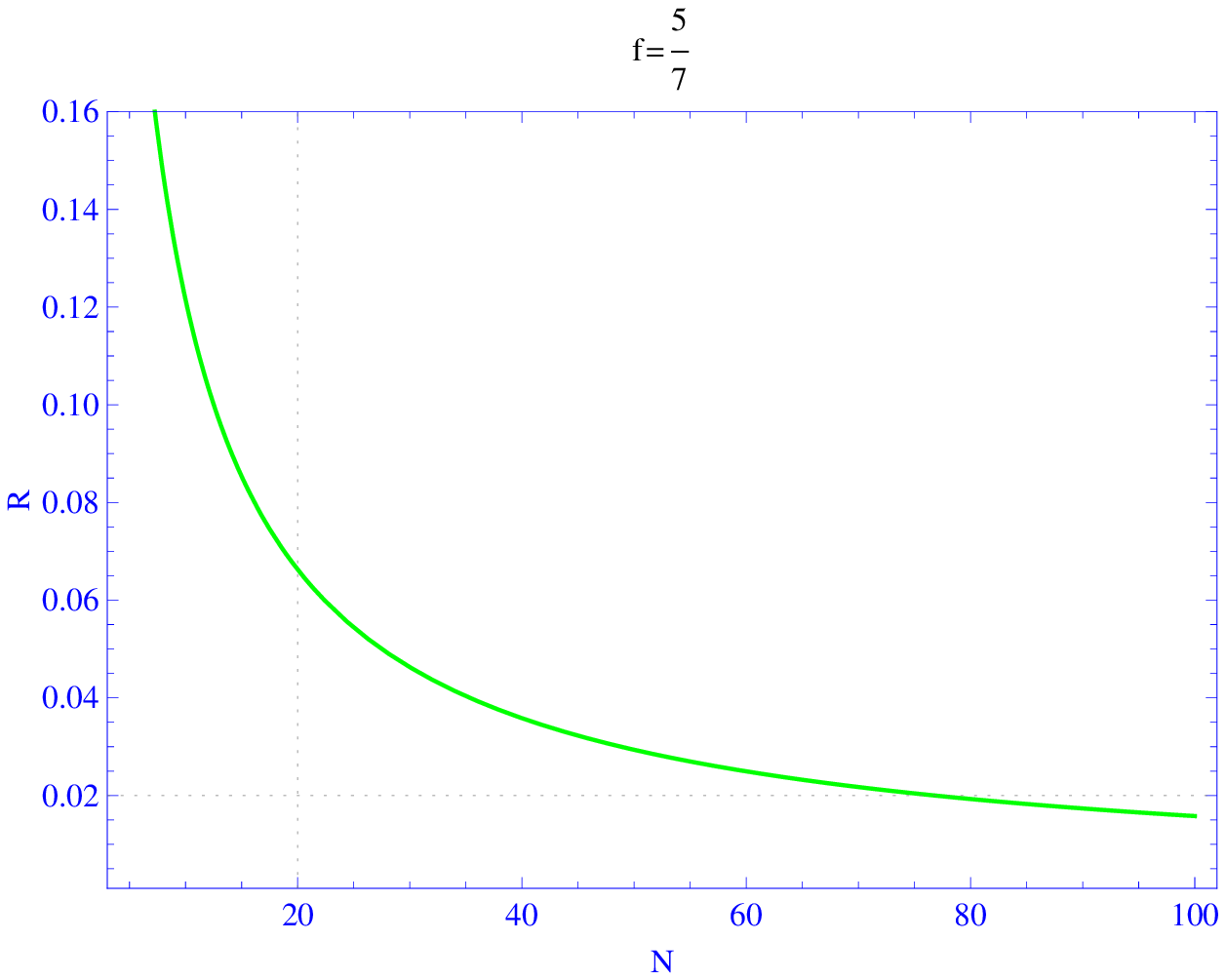}
  \caption{In this graph we plot the scalar-tensor ratio $R$ in term of the number of e-folds $N$. We can find best range of tensor-scalar ratio where $60<N<80$.}
 \label{fig:R-N-int-const}
\end{figure}

Tensor-scalar ratio of the model in this case is presented by using Eqs.(\ref{45}) and (\ref{1-i})
\begin{eqnarray}\label{}
R=B_1(\frac{\phi-\phi_0}{\beta})^{\frac{-4f+1}{2f}}\coth[\frac{k}{2T}]\\
\nonumber
=B_1(\frac{N}{A}+\frac{1-f}{fA})^{\frac{-4f+1}{4f}}\coth[\frac{k}{2T}]=B_1(\frac{4fA}{11-14f}(1-n_s))^{\frac{4f-1}{4f}}
\end{eqnarray}
where $B_1=\frac{2^{\frac{3}{2}}(4\pi)^{\frac{23}{8}}\Gamma_0^{\frac{1}{2}}\sigma^{\frac{1}{4}}(1-f)^{\frac{3}{4}}}{3^{\frac{15}{8}}M_4^{\frac{31}{4}}\lambda^{\frac{15}{8}}(fA)^{\frac{3}{2}}}(\frac{3fA}{2\lambda})^{\frac{1}{2}}$
In Fig.(\ref{fig:R-N-int-const}), tensor-scalar ratio in terms of number of e-folds is plotted where $f=\frac{5}{6}$. We could see, $60<N<80$ lead to
 $R<0.11$ \cite{planck}.
\begin{figure}[h]
\centering
  \includegraphics[width=10cm]{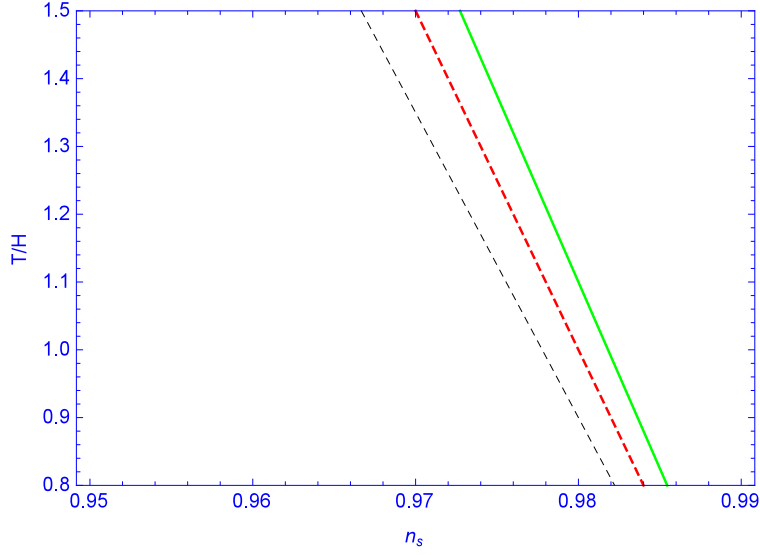}
  \caption{In this graph we plot the temperature to Huable parameter ratio $\frac{T}{H}$ in term of the spectral index $n_s$. We can find best fit of warm inflation condition ($T>H$) with the Planck data.}
 \label{fig:T-H}
\end{figure}
\begin{figure}[h]
\centering
  \includegraphics[width=10cm]{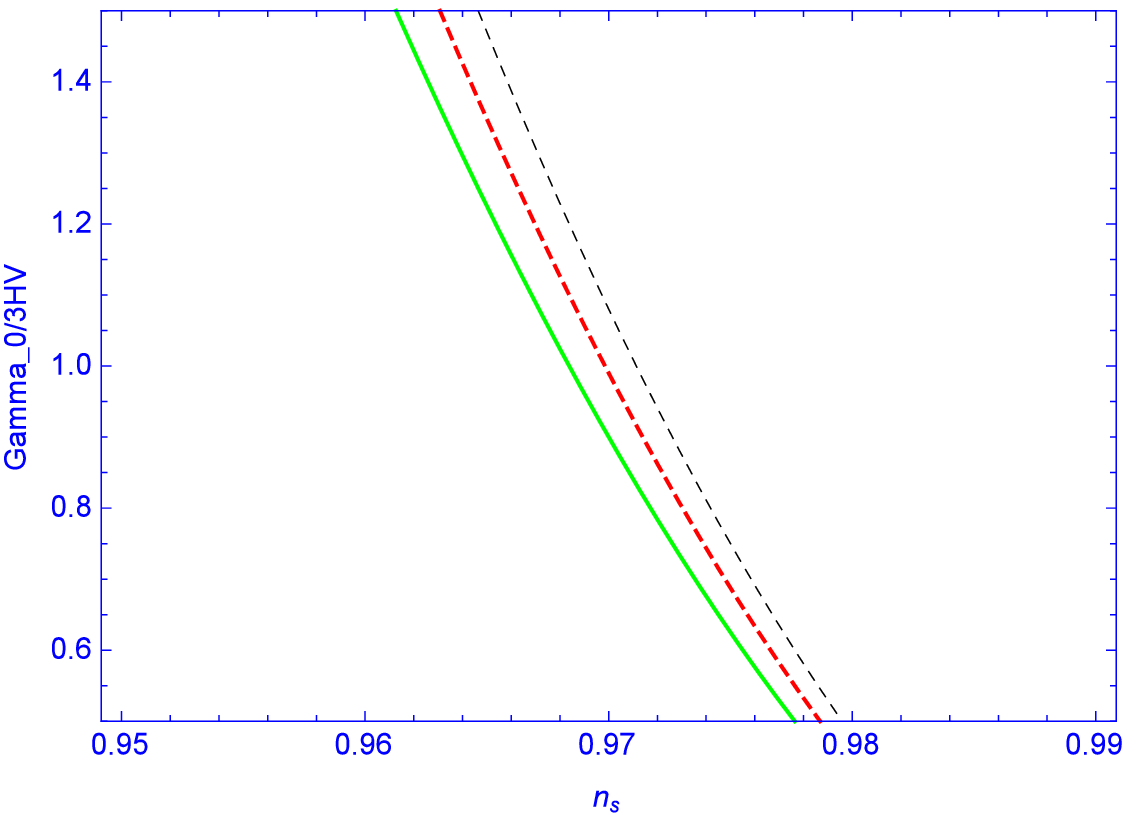}
  \caption{In this graph we plot the  dissipative to Huable parameter ratio $\frac{\Gamma_0}{3HV}$ in term of the spectral index $n_s$. We can find best fit of high dissipative regime $\Gamma_0>3HV$ with the Planck data for three cases of $\Gamma_0$.}
 \label{fig:Gamma-0}
\end{figure}
The expression for the perturbation $\delta\phi$ given by Eq.(36) is valid when $T > H$. The
the choice of the parameters of the model have to be consistent with this condition  $T > H$. In Fig. (\ref{fig:T-H}) we plot $\frac{T}{H}$ in term of spectral index that shows the model is compatible with observational data in warm inflation limit $T>H$. We also checked the high  dissipative condition $\Gamma_0>3HV$ in Fig.(\ref{fig:Gamma-0}) that we can see agreement with observational data.
\subsection{$\Gamma=\Gamma_1 V(\phi)$ case}
\begin{figure}[h]
\centering
  \includegraphics[width=10cm]{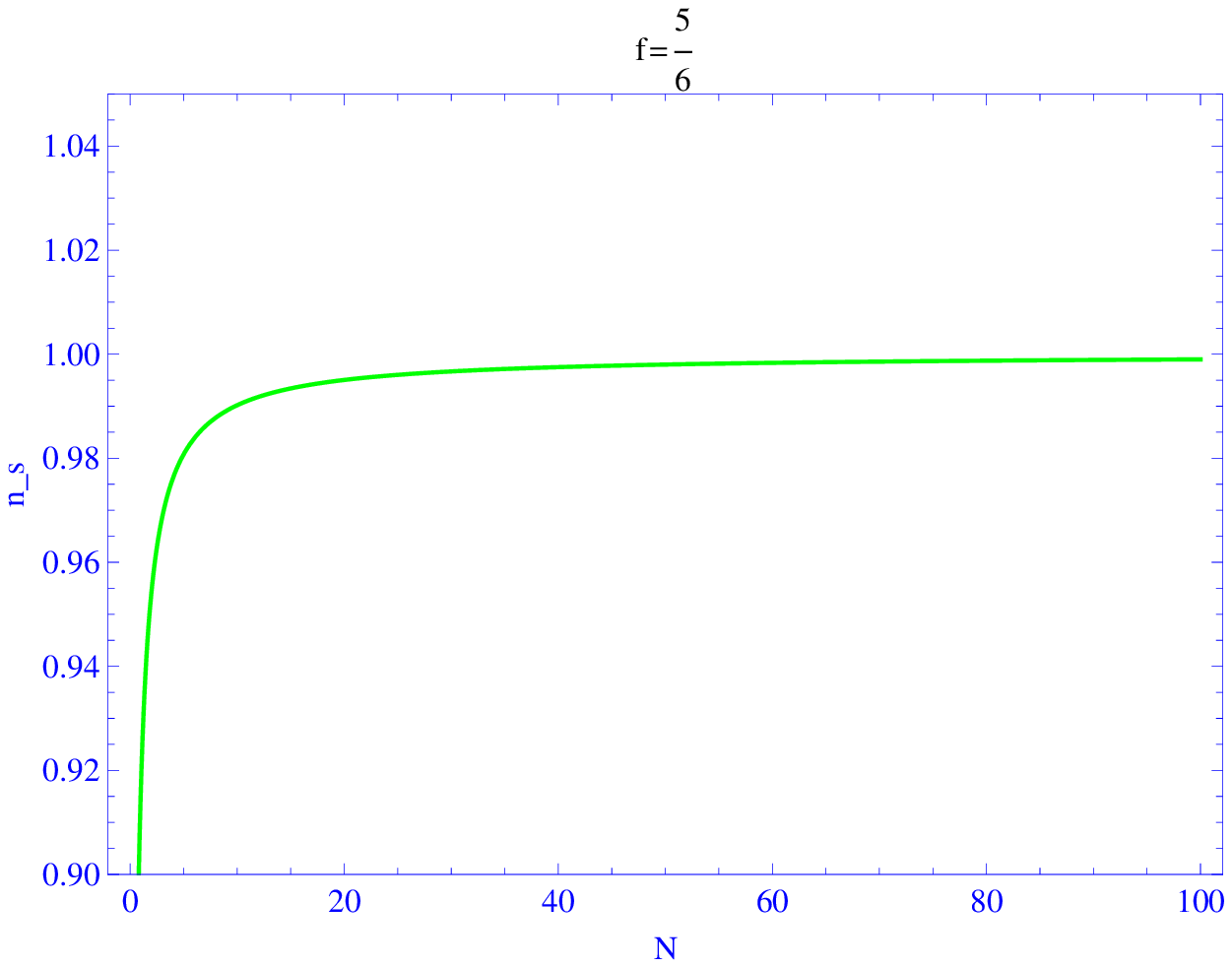}
  \caption{In this graph we plot the spectral index $n_s$ in term of the number of e-folds $N$. We can find best range of spectral index ratio where $N>60$.}
 \label{fig:n-N-int-var}
\end{figure}

\begin{figure}[h]
\centering
  \includegraphics[width=10cm]{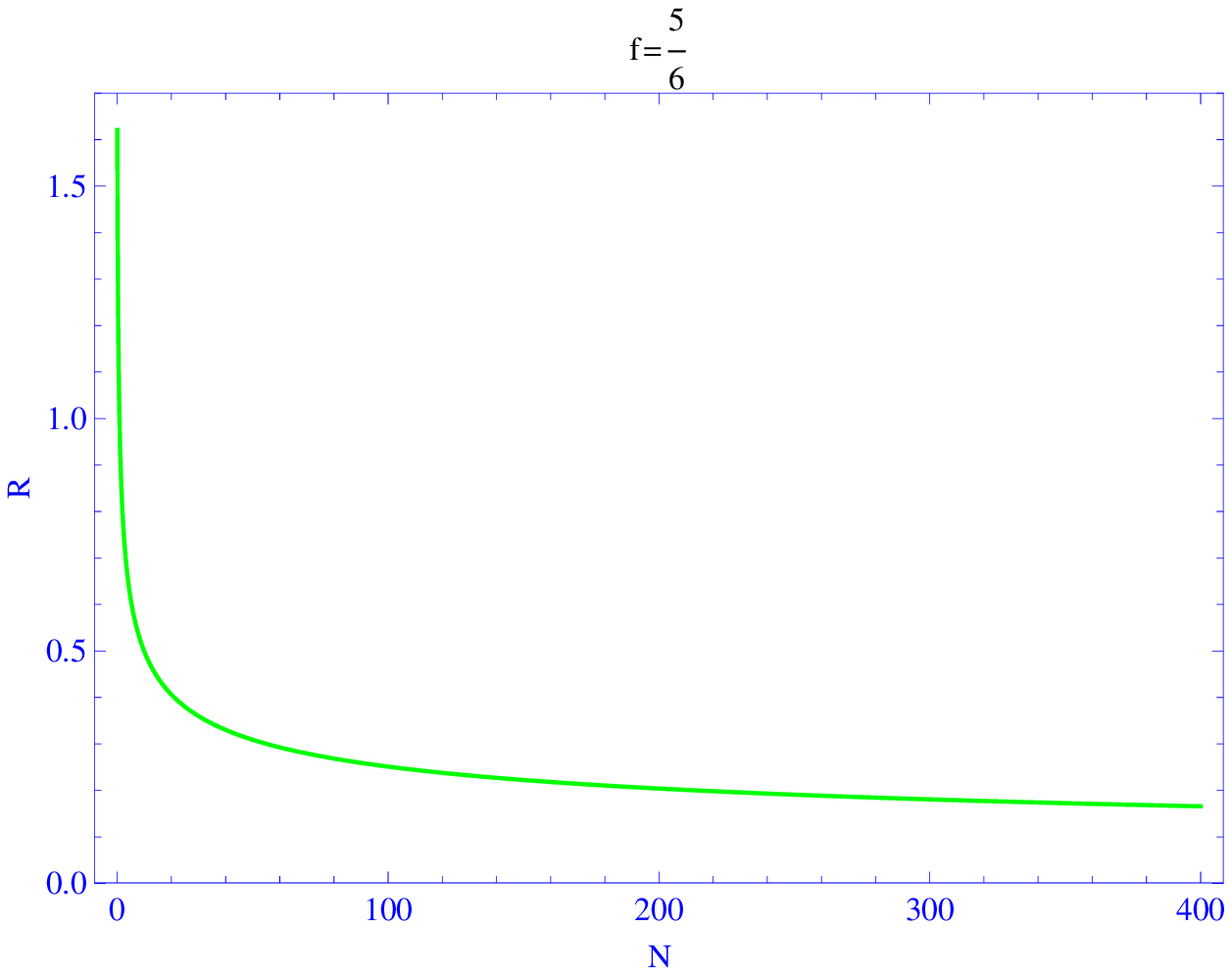}
  \caption{In this graph we plot the scalar-tensor ratio $R$ in term of the number of e-folds $N$. We can find best range of tensor-scalar ratio where $60<N$.}
 \label{fig:R-N-int-var}
\end{figure}

Dissipative parameter may be considered as a function of scalar field \cite{1-m}. We will study our model in the context of intermediate inflation where $\Gamma=\Gamma_1 V(\phi)$. In this case the scalar field is determined from Eqs. (\ref{1-i}) and (\ref{2-i})
\begin{eqnarray}\label{}
\phi-\phi_0=(\frac{4(1-f)}{\Gamma_1}t)^{\frac{1}{2}}
\end{eqnarray}
Therefor the Hubble parameter and potential of the model in terms of tachyon potential have the following forms:
\begin{eqnarray}\label{}
H(\phi)=fA(\frac{\Gamma_1(\phi-\phi_0)^2}{4(1-f)})^{f-1}~~~~~~~\\
\nonumber
V(\phi)=(\frac{3\lambda M_4^2 f^2A^2}{4\pi})^{\frac{1}{2}}(\frac{\Gamma_1(\phi-\phi_0)^2}{4(1-f)})^{f-1}
\end{eqnarray}
Dissipative parameter $r$ is presented by using above equation
\begin{eqnarray}\label{}
r=\frac{\Gamma_1 V(\phi)}{3HV}=\frac{\Gamma_1}{fA}(\frac{\Gamma_1(\phi-\phi_0)^2}{4(1-f)})^{1-f}
\end{eqnarray}
Important parameters of the slow-roll inflation in this case are presented as
\begin{eqnarray}\label{}
\epsilon=\frac{1-f}{fA}(\frac{\Gamma_1(\phi-\phi_0)^2}{4(1-f)})^{-f}\\
\nonumber
\eta=\frac{2-f}{fA}(\frac{\Gamma_1(\phi-\phi_0)^2}{4(1-f)})^{-f}
\end{eqnarray}
The number of e-folds is given by
\begin{eqnarray}\label{}
N(\phi)=A(\frac{\Gamma_1(\phi-\phi_0)^2}{4(1-f)})^{f}-A(\frac{\Gamma_1(\phi_1-\phi_0)^2}{4(1-f)})^{f}
\end{eqnarray}
where $\phi_1$ is the tachyon field at the begining of the inflation period. We find this field where the slow-roll parameter $\epsilon$ is equal to one
\begin{eqnarray}\label{}
\phi_1=\phi_0+[\frac{4(1-f)}{\Gamma_1}(\frac{1-f}{fA})^{\frac{1}{f}}]^{\frac{1}{2}}
\end{eqnarray}
From above equations we present the scalar field in term of number of e-folds and intermediate parameters $f$ and $A$
\begin{eqnarray}\label{}
\phi=\phi_0+[\frac{4(1-f)}{\Gamma_1}(\frac{N}{A}+\frac{1-f}{fA})^{\frac{1}{f}}]^{\frac{1}{2}}
\end{eqnarray}
Spectral index $n_s$ is presented using Eq.(\ref{40})
\begin{eqnarray}\label{4-i}
n_s=1+\frac{3}{4}\eta-\frac{23}{4}\epsilon=1-\frac{17-20f}{4fA}(\frac{\Gamma_1(\phi-\phi_0)^2}{4(1-f)})^{-f}\\
\nonumber
=1-\frac{17-20f}{4fA}(\frac{N}{A}+\frac{1-f}{fA})^{-1}~~~~~~~~~~~~~~~~~~~~~~~~~~~~~~~
\end{eqnarray}
We can find the scale-invariant spectrum (Harrison-Zeldovich spectrum) i.e. $n_s=1$ where $f=\frac{17}{20}$.
In Fig.(\ref{fig:n-N-int-var}),
we plot the spectral index in terms of number of e-fold where $f=\frac{5}{6}$. For $N>60$ we can see the spectral index is confined to $0.98<n_s<1$ which is compatible with Planck
 data \cite{6}.
Power-spectrum and scalar-tensor ratio of this model may be obtained from Eqs.(\ref{37}) and (\ref{45}) respectively
\begin{figure}[h]
\centering
  \includegraphics[width=10cm]{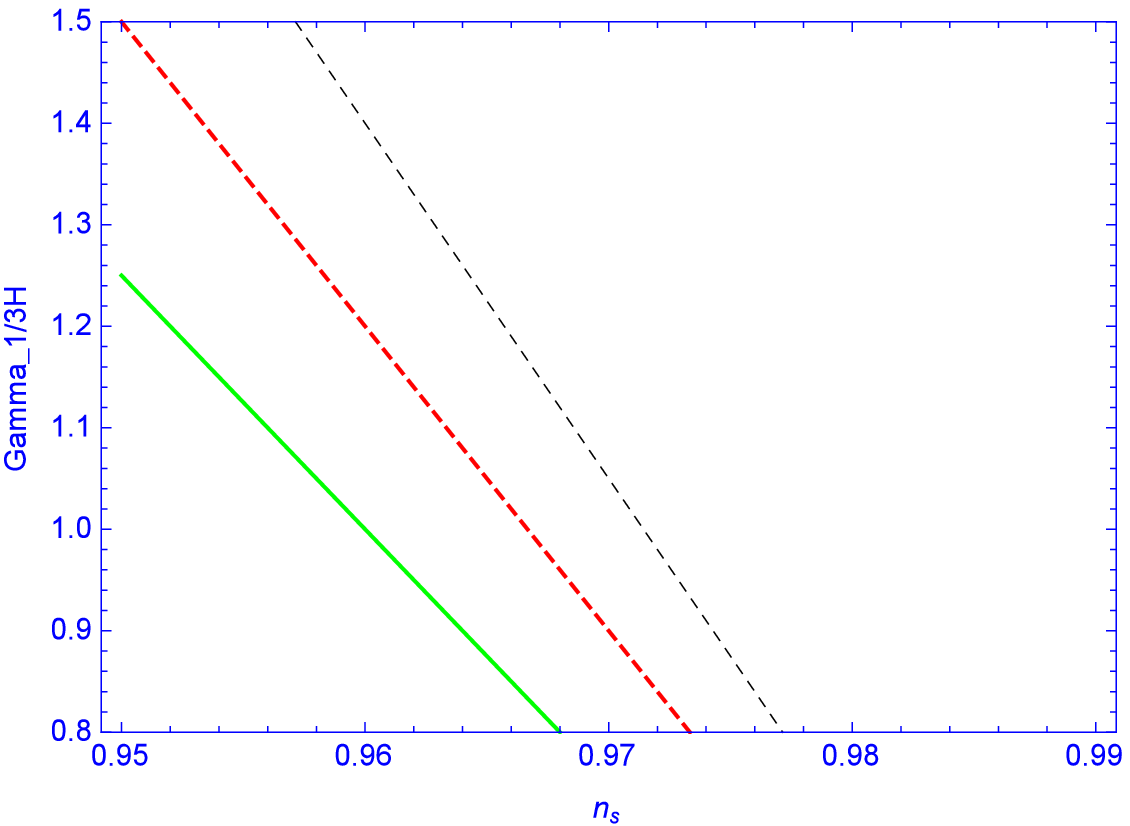}
  \caption{In this graph we plot the  dissipative to Huable parameter ratio $\frac{\Gamma_1}{3H}$ in term of the spectral index $n_s$. We can find best fit of high dissipative regime $\Gamma_0>3H$ with the Planck data for three cases of $\Gamma_1$.}
 \label{fig:Gamma-1}
\end{figure}

\begin{eqnarray}\label{3-i}
P_R=\frac{25}{4}\delta_H^2=\frac{M_4^5\lambda^{\frac{1}{2}}}{(2\pi)^{\frac{5}{2}}\sigma^{\frac{1}{4}}}\frac{V^{\frac{15}{4}}}{r^{\frac{1}{2}}\epsilon^{\frac{3}{4}}}
=A_2(\frac{\Gamma_1(\phi-\phi_0)^2}{4(1-f)})^{\frac{20f-17}{4f}}\\
\nonumber
=A_2(\frac{N}{A}+\frac{1-f}{fA})^{\frac{20f-17}{4f}}~~~~~~~~~~~~~~~~~~~~~~~~~\\
\nonumber
R=B_2(\frac{\Gamma_1(\phi-\phi_0)^2}{4(1-f)})^{\frac{-10f+7}{4f}}\coth[\frac{k}{2T}]\\
\nonumber
=B_2(\frac{N}{A}+\frac{1-f}{fA})^{\frac{-10f+7}{4f}}=B_2(\frac{fA(1-n_s)}{20f-17})^{\frac{10f-7}{4f}}\\
\end{eqnarray}
where
\begin{eqnarray}\label{}
A_2=\frac{M_4^{\frac{25}{4}}\lambda^{\frac{17}{4}}(fA)^5 3^{\frac{15}{8}}}{\sigma^{\frac{1}{4}}\Gamma_1^{\frac{1}{2}}(1-f)^{\frac{3}{4}}\pi^{\frac{35}{8}}2^{\frac{25}{4}}}~~~~~\\
\nonumber B_2=(\frac{3fA}{2\lambda})^{\frac{1}{2}}\frac{2^{\frac{39}{8}}(2\pi)^{\frac{13}{8}}\sigma^{\frac{1}{4}}\Gamma_1^{\frac{1}{2}}(1-f)^{\frac{3}{4}}}{3^{\frac{15}{8}}\lambda^{\frac{19}{8}}M_4^{\frac{35}{8}}(fA)^3}\\
\nonumber
\Im(\phi)=-\frac{9}{4}\ln(V)~~~~~~~~~~~~~
\end{eqnarray}
In Fig.(\ref{fig:Gamma-1}) we can see high dissipative condition agrees with Planck data.
In Fig.(\ref{fig:R-N-int-var}) tensor-scalar ratio in terms of number of e-folds is plotted where $f=\frac{5}{6}$. We could see, $60<N$ lead to
 $R<0.11$ \cite{planck}. 
\section{Logamediate inflation}
In this section we will study warm tachyon inflation model in the context of logamediate scenario. The scale factor of this model is given by
\begin{eqnarray}\label{1-l}
a(t)=a_0\exp(A[\ln t]^{\nu}) ~~~~~
\end{eqnarray}
where $A$ is a positive constant and $\nu>1$. We consider this model in two cases: 1- Dissipative parameter $\Gamma$ is constant. 2- Dissipative parameter is proportional to tachyon field potential $V(\phi)$.
\subsection{$\Gamma=\Gamma_0$ case}
\begin{figure}[h]
\centering
  \includegraphics[width=10cm]{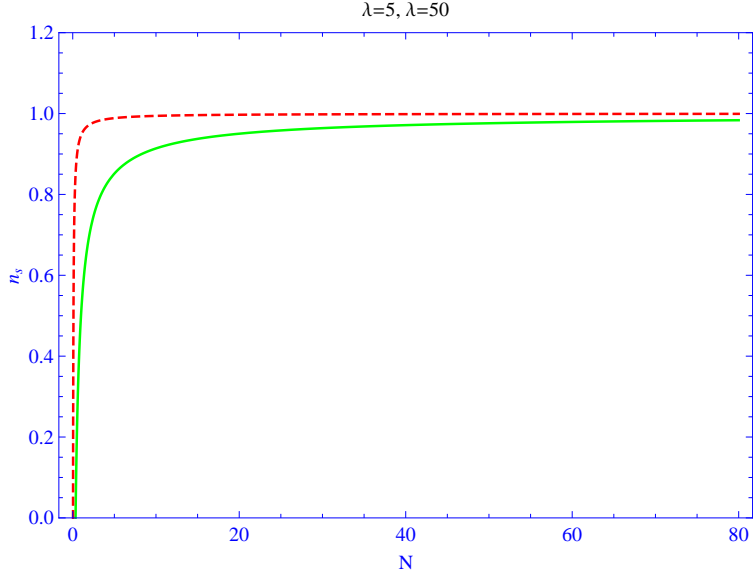}
  \caption{ Spectral index in term of number of e-folds, $\nu=50$ by dashed line and $\nu=5$ by green line. }
 \label{fig:n-N-log-const}
\end{figure}

\begin{figure}[h]
\centering
  \includegraphics[width=10cm]{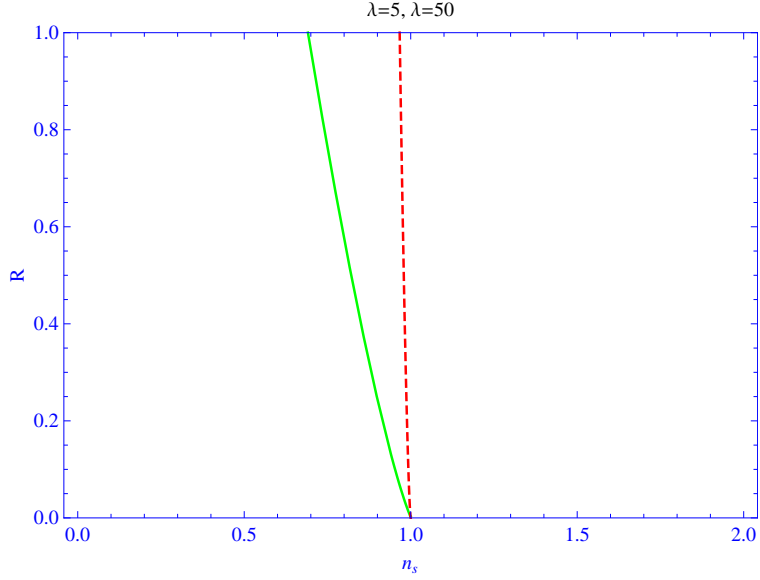}
  \caption{ Tensor-scalar ratio in term of spectral index $n_s$, $\nu=50$ by dashed line and $\nu=5$ by green line. }
 \label{fig:R-n-log-const}
\end{figure}

In this case the scalar field is given by using Eqs.(\ref{2-i}) and (\ref{1-l})
\begin{eqnarray}\label{}
\phi-\phi_0=\frac{2\omega}{\nu+1}(\ln t)^{\frac{\nu+1}{2}}
\end{eqnarray}
where $\omega=(\frac{3\lambda M_4^2 \nu^2 A^2}{2\pi \Gamma_0^2})^{\frac{1}{4}}$.
Using above equation, the Hubble parameter and tachyon potential have the following forms
\begin{eqnarray}\label{}
H=\frac{A\nu [\frac{(\nu+1)(\phi-\phi_0)}{2\omega}]^{2\frac{\nu-1}{\nu+1}}}{\exp([\frac{(\nu+1)(\phi-\phi_0)}{2\omega}]^{\frac{2}{\nu+1}})}\\
\nonumber
V=\frac{\Gamma_0\omega^2 [\frac{(\nu+1)(\phi-\phi_0)}{2\omega}]^{2\frac{\nu-1}{\nu+1}}}{\exp([\frac{(\nu+1)(\phi-\phi_0)}{2\omega}]^{\frac{2}{\nu+1}})}
\end{eqnarray}
We derive the slow-roll parameters in logamediate scenario
\begin{eqnarray}\label{}
\epsilon=\frac{1}{A\nu}[\frac{(\nu+1)(\phi-\phi_0)}{2\omega}]^{2\frac{1-\nu}{\nu+1}}\\
\nonumber
\eta=\frac{2}{A\nu}[\frac{(\nu+1)(\phi-\phi_0)}{2\omega}]^{2\frac{1-\nu}{\nu+1}}
\end{eqnarray}
The number of e-folds for present model of inflation is presented as:
\begin{eqnarray}\label{}
N=A([\ln t]^{\nu}-[\ln t_1]^{\nu})=A([\frac{(\nu+1)(\phi-\phi_0)}{2\omega}]^{\frac{2\nu}{\nu+1}}-[\frac{(\nu+1)(\phi_1-\phi_0)}{2\omega}]^{\frac{2\nu}{\nu+1}})
\end{eqnarray}
$\phi_1=\phi_0+\frac{2\omega}{\nu+1}(A\nu)^{\frac{1+\nu}{2(1-\nu)}},$ is the inflaton at the begining of the inflation era. From above equation the scalar field is presented in terms of number of e-folds
\begin{eqnarray}\label{}
\phi=\phi_0+\frac{2\omega}{\nu+1}(\frac{N}{A}+(\nu A)^{\frac{\nu}{1-\nu}})^{\frac{\nu+1}{2\nu}}
\end{eqnarray}
Dissipative parameter $r$ is given by
\begin{eqnarray}\label{}
r=\frac{\Gamma_0}{3HV}=\frac{1}{3(\nu A \omega)^2}\frac{\exp(2[\frac{(\nu+1)(\phi-\phi_0)}{2\omega}]^{\frac{2}{\nu+1}})}{[\frac{(\nu+1)(\phi-\phi_0)}{2\omega}]^{4\frac{\nu-1}{\nu+1}}}
\end{eqnarray}
Power-spectrum and scalar-tensor ratio of logamediate inflation are derived from Eqs.(\ref{37}) and (\ref{45}).
\begin{eqnarray}\label{2-l}
P_R=A_3\exp(-\frac{11}{4}[\frac{(\nu+1)(\phi-\phi_0)}{2\omega}]^{\frac{2}{\nu+1}})[\frac{(\nu+1)(\phi-\phi_0)}{2\omega}]^{7\frac{\nu-1}{\nu+1}}\\
\nonumber
=A_3\exp(-\frac{11}{4}(\frac{N}{A}+(A\nu)^{\frac{\nu}{1-\nu}})^{\frac{1}{\nu}})[\frac{N}{A}+(A\nu)^{\frac{\nu}{1-\nu}}]^{\frac{7(\nu-1)}{2\nu}}\\
\nonumber
R=B_3\exp(\frac{1}{4}[\frac{(\nu+1)(\phi-\phi_0)}{2\omega}]^{\frac{2}{\nu+1}})[\frac{(\nu+1)(\phi-\phi_0)}{2\omega}]^{4\frac{1-\nu}{\nu+1}}\\
\nonumber
=B_3\exp(\frac{1}{4}(\frac{N}{A}+(A\nu)^{\frac{\nu}{1-\nu}})^{\frac{1}{\nu}})[\frac{N}{A}+(A\nu)^{\frac{\nu}{1-\nu}}]^{\frac{4(1-\nu)}{2\nu}}
\end{eqnarray}
where
\begin{eqnarray}\label{}
A_3=\frac{M_4^5\lambda^{\frac{1}{2}}\Gamma_0^{\frac{7}{4}}\omega^{\frac{9}{2}}}{(2\pi)^{\frac{5}{2}}\sigma^{\frac{1}{2}}(\nu A)^{-\frac{7}{4}}}\\
\nonumber
B_3=(\frac{3\pi\Gamma_0\omega}{\lambda^3M_4^2})^{\frac{1}{4}}\frac{16 (2\pi)^{\frac{5}{2}}\sigma^{\frac{1}{4}}\Gamma_0^{\frac{1}{4}}}{3(3\omega)^{\frac{1}{2}}(A\nu)^{\frac{7}{4}}}\\
\nonumber
\end{eqnarray}
By using equation (\ref{40}), we could find the spectral index $n_s$
\begin{eqnarray}\label{}
n_s=1-\frac{11}{4\nu A}[\frac{(\nu+1)(\phi-\phi_0)}{2\omega}]^{2\frac{1-\nu}{\nu+1}}\\
\nonumber
=1-\frac{11}{4\nu A}[\frac{N}{A}+(A\nu)^{\frac{\nu}{1-\nu}}]^{\frac{(1-\nu)}{\nu}}
\end{eqnarray}
In Fig.(\ref{fig:n-N-log-const}), the dependence of spectral index on the number of e-folds  is shown
(for $\nu = 50$ and $\nu = 5$ cases). It is observed that the small values of the number of e-folds are assured
for large values of $\nu$ parameter. This figure shows the scale
invariant spectrum, (Harrison-Zeldovich spectrum, i.e. $n_s = 1$) could be approximately
obtained for $(\nu,N)=(50,60)$.
From above equation and Eq.(\ref{2-l}), a relation between scalar-tensor ratio and spectral index is obtained
\begin{eqnarray}\label{}
R=B_3\exp(\frac{1}{4}(\frac{4\nu A}{11}[1-n_s])^{\frac{1}{1-\nu}})[\frac{4\nu A}{11}(1-n_s)]^{2}
\end{eqnarray}
In Fig.(\ref{fig:R-n-log-const}), two trajectories in the $n_s - R$ plane are shown. There is a range of values of R
and $n_s$ which is compatible with the Planck data.
\subsection{$\Gamma=\Gamma_1V(\phi)$}
Warm tachyon inflation in the context of logmamediate scenario with dissipation $\Gamma=\Gamma_1V(\phi)$ will be studied. In this case we can find the scalar field using Eq.(\ref{2-i}) and (\ref{1-l})
\begin{eqnarray}\label{}
\phi-\phi_0=\frac{2}{\sqrt{\Gamma_1}}t^{\frac{1}{2}}
\end{eqnarray}
We also derive the Hubble parameter tachyonic potential and  dissipative parameter $r$ from above equation

\begin{eqnarray}\label{}
H(\phi)=\frac{4A\nu(\ln(\Gamma_1\frac{(\phi-\phi_0)^2}{4}))^{\nu-1} }{(\phi-\phi_0)^2}~~~~~V(\phi)=(\frac{12\lambda M_4^2A^2\nu^2}{\pi})^{\frac{1}{2}}\frac{(\ln(\Gamma_1\frac{(\phi-\phi_0)^2}{4}))^{\nu-1} }{(\phi-\phi_0)^2}~~~~~~\\
\nonumber
r=\frac{\Gamma_1}{12A\nu}\frac{(\phi-\phi_0)^2 }{(\ln(\Gamma_1\frac{(\phi-\phi_0)^2}{4}))^{\nu-1}}~~~~~~~~~~~~~~~~~~~~~~
\end{eqnarray}

\begin{figure}[h]
\centering
  \includegraphics[width=10cm]{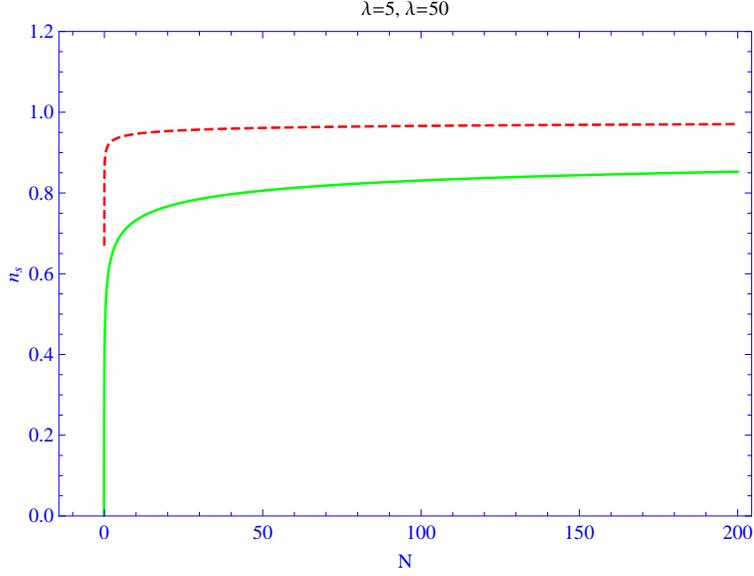}
  \caption{ Spectral index in term of number of e-folds, $\nu=50$ by dashed line and $\nu=5$ by green line. }
 \label{fig:n-N-log-var}
\end{figure}

\begin{figure}[h]
\centering
  \includegraphics[width=10cm]{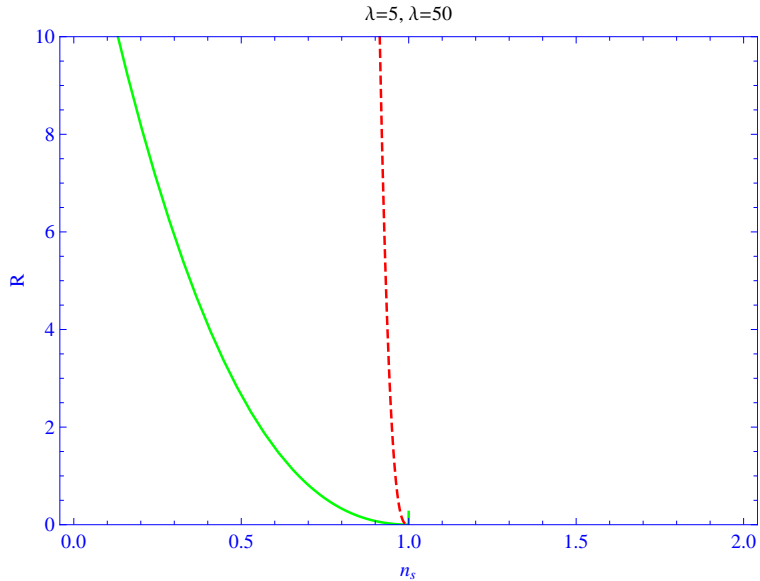}
  \caption{ Tensor-scalar ratio in term of spectral index $n_s$, $\nu=50$ by dashed line and $\nu=5$ by green line. }
 \label{fig:R-n-log-var}
\end{figure}

\begin{figure}[h]
\centering
  \includegraphics[width=10cm]{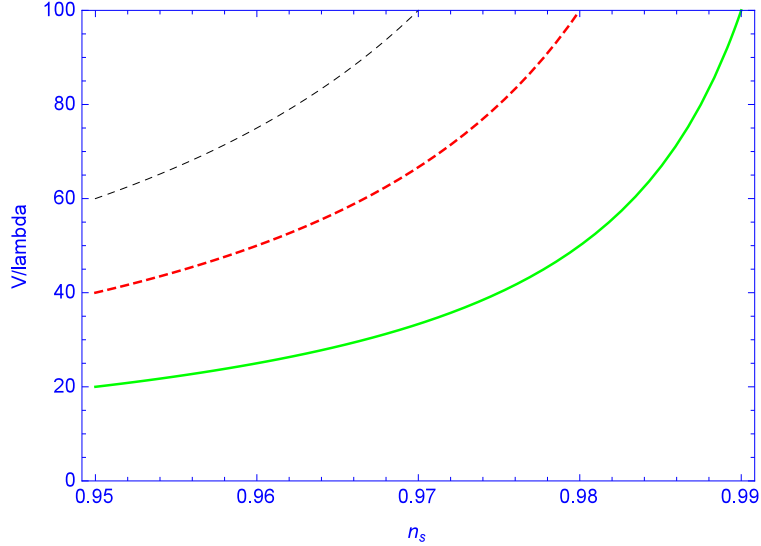}
  \caption{In this graph we plot the  tachyonic potential  in term of the spectral index $n_s$. We can find best fit of high energy limit $V\gg\lambda$ with the Planck data.}
 \label{fig:V}
\end{figure}

The slow-roll parameters $\epsilon$ and $\eta$ are presented respectively
\begin{eqnarray}\label{}
\epsilon=\frac{(\ln(\Gamma_1\frac{(\phi-\phi_0)^2}{4}))^{1-\nu}}{A\nu}\\
\nonumber
\eta=\frac{2(\ln(\Gamma_1\frac{(\phi-\phi_0)^2}{4}))^{1-\nu}}{A\nu}
\end{eqnarray}
Number of e-folds at the end of inflation is given by
\begin{eqnarray}\label{}
N=A[(\ln(\Gamma_1\frac{(\phi-\phi_0)^2}{4}))^{\nu}-(\ln(\Gamma_1\frac{(\phi_1-\phi_0)^2}{4}))^{\nu}]
\end{eqnarray}
where  $\phi_1$ is begining inflaton. At the begining point of inflaton period we have $\epsilon=1,$ therefore the inflaton in this point has the following form:
\begin{eqnarray}\label{}
\phi_1=\phi_0+\frac{2}{\sqrt{\Gamma_1}}\exp(\frac{1}{2}(A\nu)^{\frac{\nu}{1-\nu}})
\end{eqnarray}
Using above equation we could find the scalar field in terms of number of e-folds
\begin{eqnarray}\label{}
\phi_1=\phi_0+\frac{2}{\sqrt{\Gamma_1}}\exp(\frac{1}{2}[(A\nu)^{\frac{\nu}{1-\nu}}+\frac{N}{A}]^{\frac{1}{\nu}})
\end{eqnarray}
Important perturbation parameters $P_R$ (power-spectrum) and $R$ (scalar-tensor ratio) could be derived in terms of scalar field and number of e-folds
\begin{eqnarray}\label{3-l}
P_{R}=A_4(\phi-\phi_0)^{-\frac{17}{2}}[\ln(\Gamma_1\frac{(\phi-\phi_0)^2}{4})]^{\frac{20\nu-9}{4}}\\
\nonumber
=A_4(\frac{\sqrt{\Gamma_1}}{2})^{\frac{17}{2}}\exp(-\frac{17}{4}[\frac{N}{A}+(A\nu)^{\frac{\nu}{1-\nu}}]^{\frac{1}{\nu}})[\frac{N}{A}+(A\nu)^{\frac{\nu}{1-\nu}}]^{\frac{20\nu-9}{4\nu}}\\
\nonumber
R=B_4(\phi-\phi_0)^{\frac{7}{2}}[\ln(\Gamma_1\frac{(\phi-\phi_0)^2}{4})]^{\frac{-5\nu+5}{2}}\\
\nonumber
=B_4(\frac{2}{\sqrt{\Gamma_1}})^{\frac{7}{2}}\exp(\frac{7}{4}[\frac{N}{A}+(A\nu)^{\frac{\nu}{1-\nu}}]^{\frac{1}{\nu}})[\frac{N}{A}+(A\nu)^{\frac{\nu}{1-\nu}}]^{\frac{-5\nu+5}{2\nu}}
\end{eqnarray}
where
\begin{eqnarray}\label{}
A_4=\frac{3^{\frac{19}{8}}4^{\frac{9}{8}}M_4^{\frac{35}{4}}(A\nu)^5}{\pi^{\frac{35}{8}}\sigma^{\frac{1}{4}}\lambda^{-\frac{19}{8}}}\\
\nonumber
B_4=\frac{4^{\frac{1}{4}}\pi^{\frac{27}{8}}\sigma^{\frac{1}{4}}(A\nu)^{-3}}{3^{\frac{19}{8}}M_4^{\frac{35}{4}}\Gamma_1^{\frac{1}{2}}\lambda^{\frac{19}{8}}}(\frac{36A^2\nu^2}{\pi\lambda^2})^{\frac{1}{4}}
\end{eqnarray}
The spectral index $n_s$ is derived in this case as:
\begin{eqnarray}\label{}
n_s=1-\frac{17}{4A\nu}(\ln[\frac{\Gamma_1(\phi-\phi_0)^2}{4}])=1-\frac{17}{8A\nu}[\frac{N}{A}+(\nu A)^{\frac{\nu}{1-\nu}}]^{\frac{1}{\nu}}
\end{eqnarray}
In Fig.(\ref{fig:n-N-log-var}), the dependence of spectral index on the number of e-folds  is shown
(for $\nu = 50$ and $\nu = 5$ cases). It is observed that the small values of number of e-folds are assured
for large values of $\nu$ parameter. This figure shows the scale
invariant spectrum, (Harrison-Zeldovich spectrum, i.e. $n_s = 1$) could be approximately
obtained for $(\nu,N)=(50,60)$.
From above equation and Eq.(\ref{3-l}), we find the tensor-scalar ratio in term of spectral index
\begin{eqnarray}\label{}
R(n_s)=B_4(\frac{4}{\Gamma_1})^{\frac{7}{4}}\exp(\frac{7}{4}[\frac{4A\nu}{17}(1-n_s)]^{\frac{1}{1-\nu}})[\frac{4A\nu}{17}(1-n_s)]^{\frac{5}{2}}
\end{eqnarray}
In Fig.(\ref{fig:R-n-log-var}), two trajectories in the $n_s - R$ plane are shown. There is a range of values of R
and $n_s$ which is compatible with the Planck data.
In order to produce our plots, we assume some values for the several parameters ($f,A,\nu,\lambda,\Gamma_0,\Gamma_1$)
for the above cases studied, these parameters coincides within  $1\sigma$ confidence level of Planck data. We will use a new method to constrain the parameters of the model in future works. In Fig.(\ref{fig:V}) we plot the  tachyonic potential  in term of the spectral index $n_s$ in logamediate case . We can find the best fit of high energy limit $V\gg\lambda$ with the Planck data that we have used in this paper. 
\section{appendix}
In this paper we have studied the model in natural unit ($\frac{h}{2\pi}=c=1$) therefore we have ($[mass]=M$, $[time]=T$ and $[length]=L$ where $[A]$ means dimension of "$A$")
\begin{eqnarray}\label{}
[c]=LT^{-1}=1~~~~~~~~~~~~~[h]=M L^2 T^{-1}\\
\nonumber
\Rightarrow~~~~T=L=M^{-1}~~~~~~~~~~~~~~~~
\end{eqnarray}
Using Eq.(\ref{fri1}) we have
\begin{eqnarray}\label{}
[H^2]=[\frac{8\pi}{M_4^2}\rho_T(1+\frac{\rho_T}{2\lambda})]~~~~~~~~~~~~~~~~~~~~~~~\\
\nonumber
\Rightarrow \frac{[a^2]}{a^2 T^2}=\frac{[\rho_T]}{[M_4^2]}\Rightarrow~~~[\rho_T]=[T_{\mu\nu}]=[V]=[P]=M^4
\end{eqnarray}
where $V$ and $P$ are potential and pressure with dimension $M^4$.
From Eq.(\ref{rho}) we have 
\begin{eqnarray}\label{}
[\dot{\phi}]=1~~~\Rightarrow~~~~[\phi]=M^{-1}
\end{eqnarray}
It is appear tachyon scalar field has dimension $M^{-1}$ which is agree with the tachyonic potential (\ref{46}). In Eq.(\ref{5}) r.h.s and l.h.s have dimension $M^4$
\begin{eqnarray}\label{}
[\dot{\rho}]+[3H\rho]+[3HP]=[\Gamma \dot{\phi}^2]\\
\nonumber
\Rightarrow~~\frac{[\rho]}{T}+\frac{[\rho]}{T}+\frac{[P]}{T}=[\Gamma]\\
\nonumber ~~\Rightarrow [\Gamma]=M^{5}~~~~~~~~~~~~~~~
\end{eqnarray} 
In Eq.(\ref{8}) we have used dimensionless parameter $r=\frac{\Gamma}{V}\frac{1}{3H}$
\begin{eqnarray}\label{}
[r]=\frac{[\Gamma]}{[H][V]}=\frac{M^5}{M M^4}=1
\end{eqnarray}
 $\frac{V}{\Gamma}$ has dimension time ($H^{-1}$), therefore in our paper we have used $\frac{\Gamma}{V}$ instead of $\Gamma$.
 We note that from above discussion that $\chi$ in Eq.(\ref{28}) has dimension $M^{-2}$ which leads to $[C]=M^{-2}$ in Eq.(\ref{32}) and Eq.(\ref{35}) has correct dimension
 \begin{eqnarray}\label{}
[\delta\phi]=[C]\frac{[V']}{[V]}\\
\nonumber
M^{-1}=M^{-2}\frac{1}{M^{-1}}
\end{eqnarray}
In Eq.(\ref{34}), we have $2H+\frac{\Gamma}{V}$ where the analyse of dimension is given by
\begin{eqnarray}\label{}
[2H]+\frac{[\Gamma]}{V}=M+\frac{M^5}{M^4}
\end{eqnarray}
Eq.(\ref{35}) have correct dimension, for cold inflation we have $[\delta_H]=\frac{[H]}{[\dot{\phi}]}[\delta\phi]=1$, in warm inflation also we have from Eq.(\ref{35})
\begin{eqnarray}\label{}
\delta_H=[M_4^2]\frac{[V][\delta\phi]}{[V']}=M^2M^{-1}M^{-1}=1
\end{eqnarray}
We note that Eq.(\ref{36}) is in momentum space \cite{5,11-m}, Hence, inserting (\ref{36}) into (\ref{35}) means that (\ref{37}) and the following equations are in momentum space. 
\section{Conclusion and discussion }
Tachyon  inflation model on the brane with overlasting form of potential
$V(\phi)=V_0\exp(-\alpha\phi)$ which agrees with tachyon
potential properties has been studied. The main problem of the
inflation theory is how to attach the universe to the end of the
inflation period. One of the solutions of this problem is the
study of inflation in the context of warm inflation \cite{3}. In
this scenario radiation is produced during inflation period where its
energy density is kept nearly constant. This is phenomenologically
fulfilled by introducing the dissipation term $\Gamma$. The study
of warm inflation model as a mechanism that gives an end for the
tachyon inflation are motivated us to consider the warm tachyon
inflation model.
We note that, the $\Im(\phi)$ factor (\ref{34})  which is appear  in the perturbation parameters (\ref{37}), (\ref{40}), (\ref{42}) and (\ref{45}) in high energy limit ($V\gg\lambda$),  for warm tachyon inflation model on the brane has an important difference with the same factor which was obtained for usual warm tachyon inflation model \cite{1-m}
\begin{eqnarray}\label{ne}
\nonumber
\Im(\phi)=-\int[\frac{(\frac{\Gamma}{V})'}{3H+\frac{\Gamma}{V}}+(\frac{9}{8}\frac{2H+\frac{\Gamma}{V}}{(3H+\frac{\Gamma}{V})^2}~~~~~~~~~~~~~~~~~~~~~~\\
\nonumber
\times(\Gamma+4HV-\frac{\Gamma'(\ln V)'}{12H(3H+\frac{\Gamma}{V})})\frac{(\ln V)'}{V})]d\phi~~~~~~~~~~~~
\end{eqnarray}
The density square term in the effective  Einstein equation on the brane is responsible for this difference.
Therefore, the perturbation parameters which may be constrained by Planck observational data, are modified due to the effect of density square term in effective Einstein equation. Also the slow-roll parameters (\ref{10}) and (\ref{11}) which are derived in the background level, are modified because of the density square term in modified Friedmann equation (\ref{7}). The slow-roll parameters are appeared in the perturbation parameters (\ref{37}), (\ref{40}), (\ref{42}), (\ref{44}) and (\ref{45}). As have been shown in Ref.\cite{1-m} the slow-roll parameters of warm tachyon inflation model have the forms
\begin{eqnarray}\label{}
\epsilon=\frac{M_4^2}{16\pi}\frac{1}{1+r}[\frac{V'}{V}]^2\frac{1}{V}~~~~~~~~~~\\
\nonumber
\eta=\frac{M_4^2}{8\pi(1+r)V}[\frac{V''}{V}-\frac{1}{2}(\frac{V'}{V})^2]
\end{eqnarray}
These parameters are obviously different from the slow-roll parameters (\ref{10}) and (\ref{11}). Perturbation parameters of warm tachyon inflation model have following from \cite{1-m}

\begin{eqnarray}\label{}
\delta_H=\frac{\sqrt{3}}{75\pi^2}\frac{\exp(-2\Im(\phi))}{r^{\frac{1}{2}}\tilde{\epsilon}}
\end{eqnarray}

 \begin{eqnarray}\label{}
n_s=1-[\frac{3\tilde{\eta}}{2}+\tilde{\epsilon}(\frac{2V'}{V}[2\tilde{\Im}'(\phi)-\frac{r'}{4r}]-\frac{5}{2})]
\end{eqnarray}

\begin{eqnarray}\label{}
\alpha_s=\frac{2V}{V'}\tilde{\epsilon}n_s'
\end{eqnarray}

\begin{eqnarray}\label{}
n_g=-2\epsilon
\end{eqnarray}

\begin{eqnarray}\label{}
R(k_0)=\frac{240\sqrt{3}}{25m_p^2}[\frac{r^{\frac{1}{2}}\tilde{\epsilon}H^3}{T_r}\exp(2\tilde{\Im}(\phi))\coth[\frac{k}{2T}]]|_{k=k_0}
\end{eqnarray}
The above parameters are also different from the perturbation  parameters of our model on the brane  (\ref{37}), (\ref{40}), (\ref{42}), (\ref{44}) and (\ref{45}) because of the density square term in the effective  Einstein equation on the brane.
 So, from above discussion, we know the density square term in the effective Einstein equation on the brane give the significant contributions to the observable parameters, $P_R$, $R$, $n_s$ and $\alpha_s$. Also, the different observable perturbation parameters for the models of non-tachyon warm inflation and non-tachyon warm inflation model on the brane are presented in Refs.\cite{9-f} and \cite{6-f} respectively.

 In tachyon Randall-Sundrum brane-world scenario  Einstein's equation and therefore the  Friedmann equation are modified. Warm tachyon inflation parameters on the brane have  important differences with the same parameters  which were presented for usual warm inflation model \cite{6-f} because of this modification. The density square term in the effective  Einstein equation on the brane is responsible for this difference.
Therefore, the perturbation parameters which may be constrained by  Planck observational data, are modified due to the effect of density square term in effective Einstein equation and modification of tachyonic scalar field equation of motion (E.M.O) instead of normal scalar fields E.M.O.
In this article we have considered warm-tachyon
inflationary universe  model on the brane. In the slow-roll approximation the general relation between energy density of radiation and energy density of tachyon field are presented. In the longitudinal gauge and the slow-roll limit the explicit expressions
for the tensor-scalar ratio $R$ scalar spectrum $P_R$ index,
$n_s$ and its running $\alpha_s$ have been presented. We have
developed our specific model by exponential potential with a constant dissipation coefficient. In this
case we have found perturbation parameters and constrained these
parameters  Planck observational data. Intermediate  and logamediate inflation are considered for two cases of dissipative parameters: 1- $\Gamma$ is constant parameter. 2- $ \Gamma$ as a function of tachyon field. In these two cases we have found that the  model are compatible  with observational data.
Harrison-Zeldovich spectrum, i.e. $n_s = 1$ is obtained exactly by one parameter in intermediate Scenario ($f=\frac{11}{14}$ for $\Gamma=\Gamma_0$ case and $f=\frac{17}{20}$ for $\Gamma=\Gamma(\phi)$). In logamediate Scenario we have presented approximately scale-invariant  spectrum i.e. $n\simeq 1$ where $(N,\nu)=(60.50)$.


\end{document}